\documentclass[a4paper,12pt,reqno]{amsart}
\usepackage[utf8]{inputenc}
\usepackage{graphicx}
\usepackage{color}
\usepackage{array}
\newcolumntype{C}[1]{>{\centering\let\newline\\\arraybackslash\hspace{0pt}}m{#1}}

\usepackage[foot]{amsaddr}

\usepackage{subcaption}

\usepackage[usenames,dvipsnames]{xcolor}

\usepackage{tikz}
\usetikzlibrary{shapes,arrows,positioning}
\usetikzlibrary{decorations.markings}
\usetikzlibrary{decorations.pathreplacing}
\usetikzlibrary{decorations.pathmorphing,backgrounds,fit,petri}

\usepackage{amsmath}
\usepackage{amsfonts}
\usepackage{amssymb}
\usepackage{amsthm}
\usepackage{cite}
\usepackage{float}
\usepackage{changepage}
\usepackage{nicefrac}

\title[Modelling Tumor Neoneurogenesis]{Tumor-induced neoneurogenesis \\and perineural tumor growth: \\a mathematical approach}

\author[A. Bianchi]{Arianna Bianchi}
\address[A.B.]{Department of Mathematics and Maxwell Institute for Mathematical Sciences, Heriot-Watt University, Edinburgh, Scotland, EH14 4AS, UK}

\author[K. Syrigos]{Konstantinos Syrigos}
\address[K.S.]{Oncology Unit, 3rd Department of Internal Medicine, Sotiria General Hospital, Athens School of Medicine, Athens, Greece}

\author[G. Lolas]{Georgios Lolas}
\address[G.L.]{Center for Advancing Electronics Dresden, Technische Universit{\"a}t Dresden, Barkhausen Building II/7b, 01062 Dresden, Germany}

\email{ab584@hw.ac.uk , ksyrigos@med.uoa.gr , georgioslolas@gmail.com}

\date{2015}

\begin{document}

\begin{abstract}
Primary tumors infrequently lead to demise of cancer patients; instead, mortality and a significant degree of morbidity result from the growth of secondary tumors in distant organs (metastasis). It is well-known that malignant tumors induce the formation of a lymphatic and a blood vascular network around themselves. A similar but far less studied process occurs in relation to the nervous system and is referred to as \emph{neoneurogenesis}; in fact, recent studies have demonstrated that tumors initiate their own innervation. However, the relationship between tumor progression and the 
nervous system is still poorly understood. This process is most likely regulated by a multitude of factors in the tumor-nerve microenvironment and it is therefore important to study the interactions between the nervous system and tumor cells through mathematical/computational modelling: this may reveal the most significant factors of the plethora of interacting elements regulating neoneurogenesis. The present work is a first attempt to model the neurobiological aspect of cancer development through a (simple) system of differential equations. The model confirms the experimental observations that a tumor is able to promote nerve formation/elongation around itself, and that high levels of nerve growth factor (NGF) and axon guidance molecules (AGMs) are recorded in the presence of a tumor. Our results also reflect the observation that high stress levels (represented by higher norepinephrine release by sympathetic nerves) contribute to tumor development and spread, indicating a mutually beneficial relationship between tumor cells and neurons. The model predictions suggest novel therapeutic strategies, aimed at blocking the stress effects on tumor growth and dissemination.

\subsection*{Significance Statement}
The major strength of the article resides in its novelty: it constitutes a first attempt to model tumor-induced neoneurogenesis through a set of differential equations. The mathematical model presented is simple; nevertheless, it gives some interesting insights of the process. In particular, it shows that a tumor is able to induce high nerve density around itself and cause higher levels of NGF and AGMs in the tissues. Furthermore, the model confirms the empirical observation that stress (correlated with increased release of norepinephrine) affects tumor progression. The model suggests that cancer-induced neoneurogenesis is a decisive component of tumor progression and metastasis. In this regard, we envision that tumor-nerve interactions could represent a therapeutic target of primary significance. It is hoped this work will inspire further theoretical investigation into tumor neoneurogenesis. Also, we hope that experiments will follow to investigate proposed hypotheses such as the inverse dependence of tumor growth and progression on stress modulators (through an Allee effect).

\end{abstract}

\maketitle

\section{Introduction}
A relationship between tumors and the nervous system has been suspected since the second 
century AD with the work of the Greek physician Galen \cite{thaker2008}. Traditionally, 
the nervous system has not been considered to be actively involved in the process of metastasis. 
However, recent studies have demonstrated the presence of neurons in peritumoral regions of several 
human tumors, and the number of tumor-associated neurons has been correlated with metastases 
\cite{ayala2008,lu2003}. The relative importance of pre-existing versus newly-formed neurons 
to metastasis is not understood. Although pre-existing peritumoral neurons are likely to be 
sufficient for tumor spread, recruitment of neurons into the close proximity of a tumor may 
increase the propensity of tumors to metastasize. Increased nerve density and/or presence of 
intratumoral neurons should be regarded as an additional pathway for metastasis.

Significant progress has also been made in understanding the effects of stress- and depression-mediated release of chemicals by the nervous system on tumor cell dissemination \cite{vissoci2004, antoni2006,lutgendorf2011}. 
On the one hand tumor cells produce factors that induce the formation of a neural network, and on the other the newly 
formed nerves release neurotransmitters that affect tumor growth and migration 
\cite{lang2007,magnon2013}. Following the terminology suggested in \cite{entschladen2006}, 
the formation of new nerve branches is herein called \emph{neoneurogenesis}, 
in analogy to lymphangiogenesis and (blood) angiogenesis.

%
%

The present model aims to investigate how solid tumors induce peripheral nerve proliferation and how different types of nerves affect tumor growth and metastasis by releasing substances such as neurotrasmitters. Also, we address the question of which role stress plays in cancer progression.
This model was mainly inspired by recent works presented in \cite{ayala2008, magnon2013}, focussing predominantly on prostate cancer \cite{syrigos2001}. 
The study in \cite{ayala2008} combines \emph{in vitro} experiments with autopsy analysis of prostate cancer patients; in \cite{magnon2013} the authors explore the effects of the nervous system on tumor progression by altering nerve structure and receptor activity in mice, after implanting human tumor cells in the animals. Since the scope of our work does not include tumorigenesis, our model simulations start with a nonzero initial condition for primary tumor cells, reflecting the tumor cells implanation described in \cite{magnon2013}. Our aim is to investigate the further evolution of these cells and their interactions with the preexisting prostate-surrounding nerves. The model takes major inspiration from \cite{ayala2008}, supporting the hypothesis of a symbiosis between nerves and tumor cells.

\section{Biological background} 

\subsection{Neurons, neurotransmitters and the Autonomic Nervous System (ANS)}
Neurons (or \emph{nerve cells}) are the core components of the nervous system. 
The electrical signals travelling inside a neuron are converted into signals trasmitted by certain chemicals (\emph{neurotransmitters}); 
these are then passed to another neuron across a \emph{synapse}. A neurotransmitter released by a nerve 
binds to a receptor on another cell and, according to the receptor type, induces a certain action.
The collection of all the neuronal structures that together control body functions below the level of consciousness (for instance, heart and respiratory rate, digestion and pupillary dilation) constitute the ANS. The ANS is in turn made of three sub-systems; 
here we will focus only on two of them: the Sympathetic Nervous System (SNS, also called ``fight or flight'' system), which is 
responsible for quick response processes, and the Parasympathetic Nervous System (PNS, also known as ``rest and digest'' system), which 
governs slower responses such as gastro-intestinal functions.

\subsection{Tumor-induced neoneurogenesis}
Tumors induce innervation around themselves \cite{entschladen2008,lu2003,palm2007} and, in general, high levels of innervation in tumors correlate with a poor disease outcome \cite{ayala2008,magnon2013}.
Tumor cells have the ability to produce substances, such as Nerve Growth Factor (NGF), that stimulate the growth and improve the survival of nerve cells  \cite{Ldolle2003,geldof1998,ricci2001}.
NGF also promotes tumor growth \cite{Ldolle2003} and inhibits aggregation of cancer cells and thus enhances tumor invasion, although this process is currently poorly understood \cite{Ldolle2005}.

Tumors also release Axon Guidance Molecules (AGMs). These molecules were originally considered only for their role in the nervous system as guidance cues for axons.\footnote{The term 
\emph{axon guidance} denotes the process by which neurons send out axons along a precise path in order to reach the correct targets. The tip of an axon 
(or \emph{growth cone}) is equipped with receptors that can sense (gradients of) chemicals, called \emph{guidance cues}, which ``tell'' them where to expand 
\cite{dickson2003}.} 
In recent years, however, it was shown that many AGMs can also influence neuronal survival and migration and likely play an important role in cancer progression \cite{mehlen2011}.
There are at least three different families of AGMs (semaphorins, slits and netrins), which seem to have different roles in nervous system development and cancer progression. They are found in many different body tissues and can regulate cell migration and apoptosis (for a review of the role of AGMs in cancers, see \cite{chedotal2005}).

\subsection{ANS effects on tumor progression}
It was originally believed that the nervous system only indirectly affected cancer development, through perineural invasion 
(that is, the spread of tumors along nerve fibers \cite{entschladen2007,villers1989}) 
and modulation of the immune function \cite{vissoci2004}.
Indeed, neurotransmitters regulate the cytotoxicity of T lymphocytes and natural killer cells \cite{lang2003} and induce leukocyte migration \cite{entschladen2002,lang2007}; the consequent immunosuppression can favor tumor growth and progression, impairing the anti-tumor response \cite{balkwill2001,vissoci2004}. 
However, it is the migratory effect of neurotransmitters that first suggested a \emph{direct} link between nerves and tumor progression. One theory for the spread of metastases from a primary tumor to a certain organ claims that circulating cancer cells are attracted and settle in a specific region of the body due to the presence of factors such as chemokines or AGMs \cite{chedotal2005,liotta2001}. This assumption is in agreement with the well-known ``seed and soil'' hypothesis \cite{fidler2003}.
In particular, several studies have shown that neurotransmitters influence the migratory activity 
of cancer cells, perhaps by inducing a phenotypic change towards a more motile phenotype via intracellular signalling \cite{entschladen2008,ondicova2010}, or simply by chemotaxis \cite{drell2003}. In addition, some neurotransmitters also induce tumor growth \cite{lang2007}. Indeed, tumor cells express many receptors, including serpentine receptors \cite{entschladen2004} to which neurotransmitters are ligands. Neurotransmitters can induce several behavioral changes in tumor cells, mostly increasing their proliferation and/or migration (a summary of such effects can be found in~\cite{lang2007}).


\section{Mathematical model}

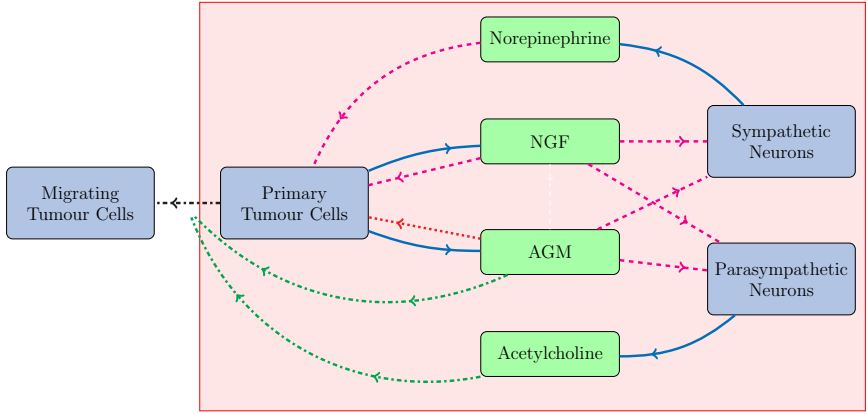
\begin{figure}[ht]
\begin{center}

\tikzstyle{cell} = [rectangle, draw, fill=NavyBlue!30,text width=7.5em, text centered, rounded corners, minimum height=4em]
\tikzstyle{chemical} = [rectangle, draw, fill=green!35,text width=7em, text centered, rounded corners, minimum height=2.5em]
\tikzstyle{line} = [draw, -latex']
\tikzset{->-/.style={decoration={
  markings,
  mark=at position .5 with {\arrow{>}}},postaction={decorate}}}
\tikzset{->--/.style={decoration={
  markings,
  mark=at position .25 with {\arrow{>}}}, postaction={decorate}}}
\tikzset{-->-/.style={decoration={
  markings,
  mark=at position .75 with {\arrow{>}}}, postaction={decorate}}}
\tikzset{->->-/.style={decoration={
  markings,
  mark=at position .30 with {\arrow{>}},
  mark=at position .75 with {\arrow{>}}}, postaction={decorate}}}

\tikzstyle{solid}=[dash pattern=]
\tikzstyle{dashed}=[dash pattern=on 3pt off 3pt]
\tikzstyle{dotted}=[dash pattern=on \pgflinewidth off 2pt]
\tikzstyle{dashdotted}=[dash pattern=on 3pt off 2pt on \the\pgflinewidth off 2pt]

\resizebox{0.9\textwidth}{!}{
\begin{tikzpicture}[node distance = 4 cm, auto] 

\filldraw[draw=red,fill=red!10!white] (-8,3.2) rectangle (7.2,-6.2);
 
\node [chemical] (ngf) {NGF};
\node [chemical, above=1.3cm of ngf] (NorAdr) {Norepinephrine};
\node [chemical, below=1.5cm of ngf] (agm) {AGM};
\node [chemical, below=1.3cm of agm] (ach) {Acetylcholine};
\node [cell, right=2cm of ngf] (symp) {Sympathetic Neurons};
\node [cell, below=1.5cm of symp] (parasymp) {Parasympathetic Neurons};

\draw[dashed,white, ->--] (ngf) -- node [midway, below, name=tumpos] {} (agm);

\node [cell, left=4cm of tumpos] (tumour) {Primary Tumour Cells};
\node [cell, left=1.5cm of tumour] (MigTum) {Migrating Tumour Cells};

\draw[ultra thick, dashdotted,Black, -->-] (tumour) -- node [midway, below, name=migr] {} (MigTum);


\draw[ultra thick, solid,NavyBlue, -->-] (tumour)  to [bend left=10]   (ngf);
\draw[ultra thick, solid,NavyBlue, -->-] (tumour) to [bend right=10] (agm);

\draw[ultra thick, solid,NavyBlue, -->-] (symp)  to [bend right=20]  (NorAdr);
\draw[ultra thick, solid,NavyBlue, -->-] (parasymp) to [bend left=20]  (ach);

\draw[ultra thick, dashed,Magenta, -->-] (NorAdr) to [bend right=30]  (tumour);

\draw[ultra thick, dashed,Magenta, -->-] (ngf) to (tumour);
\draw[ultra thick, dashed,Magenta, -->-] (ngf) to (symp);
\draw[ultra thick, dashed,Magenta, -->-] (ngf) to (parasymp);

\draw[ultra thick, dotted,Red, -->-] (agm) to (tumour);
\draw[ultra thick, dashdotted,Green, ->->-] (agm) to [bend left=35](migr);
\draw[ultra thick, dashed,Magenta, -->-] (agm) to (symp);
\draw[ultra thick, dashed,Magenta, -->-] (agm) to (parasymp);

\draw[ultra thick, dashdotted,Green, ->->-] (ach) to [bend left=40] (migr);


\end{tikzpicture}
}

\caption{A schematic representation of the interactions among the model variables. Each variable corresponds to a rounded-corners rectangular box; note that cells are in {\color{NavyBlue} \textsc{blue}} while chemicals are in {\color{Green} \textsc{green}}. The light red-shaded rectangular area represents the main domain, that is the prostate and its immediate surroundings.
Concerning the arrows, {\color{NavyBlue} \textbf{solid blue}} denotes production, 
{\color{Magenta} \textbf{dashed magenta}} denotes enhancement of growth and/or survival (and axonal extension in the case of neurons),
{\color{Green} \textbf{dash-dotted green}} denotes migration enhancement, \textbf{dash-dotted black} actual migration
and {\color{Red} \textbf{dotted red}} denotes apoptosis induction. }

\label{fig:NeuroModelScheme}

\end{center}
\end{figure}

We define the \emph{main domain} of our study as a portion of the body containing the prostate and its near surroundings, thus including both the tumor and the neighboring nerves. All the variables, with the exception of the migrating tumor cells (see below), are average concentrations/densities over this domain, which vary in time. We develop a compartmental model in which an \emph{extra domain} is considered for the tumor cells which leave the main domain. A schematic of the model, showing the variables and their interactions, can be found~in Figure \ref{fig:NeuroModelScheme}.

We distinguish between \emph{primary} tumor cells ($T_p$) and \emph{migrating} tumor cells ($T_m$).
The former are those that constitute the original tumor mass;  when they detach and leave the orthotopic 
site of the tumor they are then designated migrating. The migrating cells are particularly dangerous 
because they have the potential to form metastases. Herein we do not explicitly account for the further 
development of the migrated tumor cells: our variable $T_m$ represents an indication of \emph{potential} 
metastasis formation. 

NGF ($G$) is a neurotrophin (a kind of protein) which stimulates the growth and enhances the survival of both Sympathetic Nerve Cells (SNCs) and Parasympathetic Nerve Cells (PNCs). It has been found to be secreted by tumor cells. 
AGMs ($A$) also affect the survival and moreoever the growth of both SNCs and PNCs. In reality, there are many kinds of AGMs, which can have completely different effects on nerve and tumor development. Here, for simplicity, we consider them as a single variable; taking into account the different types of AGMs would be one step towards improving the model in future.

The growth of both SNCs ($S$) and PNCs ($P$) is enhanced by NGF and AGMs. In addition, both types of nerve cell respond to a neurotransmitter called \emph{acetylcholine} ($N_a$), but only PNCs produce it; SNCs instead secrete \emph{epinephrine} (also known as \emph{adrenaline}) and \emph{norepinephrine} ($N_n$, also called \emph{noradrenaline}).
Furthermore,  norepinephrine enhances tumor cell survival, growth and chemotaxis whereas acetylcholine seems to stimulate tumor cell invasion and migration \cite{magnon2013}.


\subsection{Model equations}


It is well documented that tumor cells naturally undergo mitosis. The model accounts for this by taking constant growth rates $r_T$ and $r_{T_m}$ for primary and migrating tumor cells, respectively. Only a fraction of primary tumor cells exhibit proliferation; this is due to the presence of a necrotic core, that we assume to be defined by the half inner radius of the (spherical) tumor mass \cite{ballangrud1999}. This assumption leads to the conclusion that only 7/8 of the tumor volume (and thus primary tumor cells) proliferate. Primary tumor cells are also exposed to the chemicals present in the domain which influence the tumor development. Since tumor growth is enhanced by NGF \cite{zhu2001}, we assume that the growth rate of $T_p$ is increased in a saturating manner by this factor. 
It has been shown that a classic logistic equation is often not suitable for modelling tumor growth \cite{korolev2014}. Here we include an Allee effect\footnote{The \emph{Allee effect} is an ecological term describing a correlation between the size and the per capita growth rate of a population.} in the growth term to take into account the fact that tumor cell populations tend to die out at low densities. This effect is traditionally found in ecological literature; its inclusion in cancer modelling was already suggested in \cite{korolev2014}. The use of ecological concepts in cancer biology and modelling is a promising development in tumor research \cite{pienta2008}. Here we take the \emph{Allee threshold}, as defined in \cite{korolev2014}, to be a function $\vartheta= \vartheta(N_n)$ that decreases as the norepinephrine level increases. This choice reflects the observation that norepinephrine enhances tumor cell survival \cite{eng2014,sood2010}.
Tumor cells also die at a constant rate $d_T$. Interestingly, some AGMs (such as netrin-1) are also thought to control tumor cell apoptosis \cite{latil2003,thiebault2003}; we model this phenomenon by adding a linear dependence on $A$ to the death term.
Finally, another relevant aspect of tumor cell dynamics is migration. Tumor cells can spontaneously disaggregate and move away from their original site \cite{pienta1995}. This process is enhanced by substances produced by nerve cells and distant organs, including AGMs \cite{capparuccia2009} and acetylcholine \cite{entschladen2002,lang2007}.
Hence, the densities of primary and migrated tumor cells are described by the following equations:
\begin{align} 
\label{eq:Tp}
\frac{dT_p}{dt}   = & \frac{7}{8}T_p \cdot \underbrace{ \left( r_T + \frac{G}{\tau_1 +\tau_2 G} \right)  }
                          _{ \mbox{\footnotesize growth upregulated by NGF} }  
            \cdot \underbrace{ \left( 1-\frac{T_p}{k_T} \right) \cdot \left( \frac{T_p}{\vartheta(N_n)} - 1\right) }
                          _{ \mbox{\footnotesize logistic growth with Allee effect} }  \nonumber \\
   & -\underbrace{d_T \left(1+\delta A\right) \cdot T_p }_{ \stackrel{\mbox{\footnotesize cell death }}{\mbox{\footnotesize increased by AGM}} } - \underbrace{ ( \mu_0 + \mu_1 A + \mu_2 N_a ) \cdot T_p}_{ \stackrel{\mbox{\footnotesize migration induced by}}{\mbox{\footnotesize AGM and acetylcholine}} }   \; , \\
\label{eq:Tm}
\frac{dT_m}{dt}   = & \underbrace{ (r_{T_m}-d_T)\cdot T_m }_{ \stackrel{\mbox{\footnotesize natural cell}}{\mbox{\footnotesize growth/death}} }
                    + \underbrace{ ( \mu_0 + \mu_1 A + \mu_2 N_a ) \cdot T_p}_{ \stackrel{\mbox{\footnotesize tumor cell migration}}{\mbox{\footnotesize from the primary tumor}} }    \; ,                      
\end{align}
where\footnote{See Appendix \ref{app-NeuroTeqn} for the motivation of the definition of $\vartheta(N_n)$.}
\begin{equation*}  
\vartheta (N_n) = \frac{\theta_1}{1+\theta_2 N_n} \; .
\end{equation*}


\noindent
We are interested in the effects that tumor-secreted NGF and AGMs have on the system; here the tumor secretion rate of these two growth factors is assumed to be constant \cite{entschladen2008,entschladen2006}. We do not include other sources of NGF and AGMs in the main domain since these have a negligible effect on the dynamics that we want to study here (their effect on nerve growth in absence of tumor is implicitly included in the logistic growth of nerve cells -- see below).
As chemicals, both NGF and AGMs decay at constant rate $d_G$ and $d_A$, respectively. They are also internalized by both tumor \cite{rakowicz1993} and nerve cells \cite{claude1982}, which bind them to their surface receptors. Here we assume that SNCs and PNCs bind the proteins at the same rate (namely, $\gamma_2$ for NGF and $\gamma_4$ for AGM).
The evolution equations describing NGF and AGM dynamics in the domain are therefore given by
\begin{align} 
\label{eq:NGF}
\frac{dG}{dt}   & = \underbrace{s_G T_p}_{ \stackrel{\mbox{\footnotesize production by}}{\mbox{\footnotesize tumor cells}} }
                      -\underbrace{d_G G}_{\mbox{\footnotesize decay}} 
                      -\underbrace{\left[ \gamma_1 T_p+\gamma_2 (S+P)\right]G}_{ \stackrel{\mbox{\footnotesize internalization by tumor}}{\mbox{\footnotesize and nerve cells}} } \; , \\
\label{eq:AGM}
\frac{dA}{dt}   & = \underbrace{s_A T_p}_{ \stackrel{\mbox{\footnotesize production by}}{\mbox{\footnotesize tumor cells}} }
                      -\underbrace{d_A A}_{\mbox{\footnotesize decay}} 
                      -\underbrace{\left[ \gamma_3 T_p+\gamma_4 (S+P)\right] A}_{ \stackrel{\mbox{\footnotesize internalization by tumor}}{\mbox{\footnotesize and nerve cells}} }   \; . 
\end{align}


\noindent
We assume that in a normal (i.e. tumor-free) setting both SNCs and PNCs grow in a logistic manner and tend to their carrying capacities $k_S$ and $k_P$, which are equal to their normal equilibrium values. However, when tumor cells are present nerve growth is enhanced by the secreted NGF \cite{lindsay1988,madduri2009} and AGMs \cite{wang1999}. This additional growth is modelled by two saturating functions and is not subject to logistic limitation. This is due to the fact that, given the complex shape of neurons, it is difficult to estimate an actual maximum density. Nerve growth can also occur as axon elongation, which does not take a significant portion of space.
Thus, the equations characterizing SNC and PNC rate of change are
\begin{align} 
\label{eq:SympNC}
\frac{dS}{dt}   & = \underbrace{ r_S \left( 1 - \frac{S}{k_S} \right) \cdot S  }
                       _{ \stackrel{\mbox{\footnotesize logistic growth}}{\mbox{\footnotesize and remodelling}} } 
					   + \underbrace{ \left( \frac{G}{\sigma_1 +\sigma_2 G} + \frac{A}{\sigma_3 +\sigma_4 A} \right) \cdot S }
                       _{ \stackrel{\mbox{\footnotesize extra growth upregulated}}{\mbox{\footnotesize by NGF and AGM}} }  
					   \; , \\
\label{eq:ParasympNC}
\frac{dP}{dt}   &  = \underbrace{ r_P \left( 1 - \frac{P}{k_P} \right) \cdot P  }
                       _{ \stackrel{\mbox{\footnotesize logistic growth}}{\mbox{\footnotesize and remodelling}} } 
                       + \underbrace{ \left( \frac{G}{\pi_1 +\pi_2 G} + \frac{A}{\pi_3 +\pi_4 A} \right) \cdot P }
                       _{ \stackrel{\mbox{\footnotesize extra growth upregulated}}{\mbox{\footnotesize by NGF and AGM}} } \; .
\end{align}


\noindent
Norepinephrine and acetylcholine are produced by SNCs and PNCs, respectively, \cite{guytonBOOK} at respective net rates $s_n$ and $s_a$  that we assume to be constant.
However, these two neurotransmitters are also released by other cell types \cite{miller2000,wessler2008} and we include constant sources $c_n$, $c_a$ in their equations.
As chemicals, they decay at constant rates $d_n$ and $d_a$, respectively.
Finally, they are absorbed by tumor cells \cite{lang2007,magnon2013} at constant rates $\gamma_5$ and $\gamma_6$, respectively.
The evolution equations for the neurotransmitters are then expressed by
\begin{align} 
\label{eq:Norep}
\frac{dN_n}{dt} & =  \underbrace{c_n}_{\stackrel{\mbox{\footnotesize const.}}{\mbox{\footnotesize source}}} + \underbrace{s_n S}_{ \stackrel{\mbox{\footnotesize production}}{\mbox{\footnotesize by SNCs}} }
                      -\underbrace{d_n N_n}_{\mbox{\footnotesize decay}} 
                      -\underbrace{\gamma_5 T_p N_n}_{ \stackrel{\mbox{\footnotesize uptake by}}{\mbox{\footnotesize tumor cells}} } \; , \\
\label{eq:Acetylch}
\frac{dN_a}{dt} & =  \underbrace{c_a}_{\stackrel{\mbox{\footnotesize const.}}{\mbox{\footnotesize source}}} + \underbrace{s_a P}_{ \stackrel{\mbox{\footnotesize production}}{\mbox{\footnotesize by PNCs}} }
                      -\underbrace{d_a N_a}_{\mbox{\footnotesize decay}}
                      -\underbrace{\gamma_6 T_p N_a}_{ \stackrel{\mbox{\footnotesize uptake by}}{\mbox{\footnotesize tumor cells}} }  \; .
\end{align}


\subsection{Parameters and initial conditions}

\subsubsection*{Parameters}
Table \ref{table:paramNeuroODEs} reports a list of all the parameters appearing in the model equations. 
Each parameter is supplied with its estimated value, units and source used (when possible) to assess it. References in brackets mean that although the parameter was not \emph{directly} estimated from a dataset, its calculated value was inspired by the biological literature. When no data were found to inform a parameter value, this was taken to be of the same order of magnitude of another reasonably similar one.
A detailed description of the estimation of each parameter can be found in Appendix \ref{appNeuroPAR}.

\begin{table}[ht]

\begin{small}

\makebox[\textwidth][c]{\begin{tabular}{cccc}
\hline
\textsc{parameter} & \textsc{value} &  \textsc{units}  &  \textsc{source}   \\
\hline
$r_T$              & $4.81 \times 10^{-4}$   &    day$^{-1}$                 & \cite{schmid1993}  \\
$r_{T_m}$          & $1 \times 10^{-4}$      &    day$^{-1}$                 & estimated$\approx r_T$  \\
$\tau_1$           &   $307.4$      & $\mbox{pg day }(\mbox{mm}^3)^{-1}$     & \cite{zhu2001} \\
$\tau_2$           &    $22.7$      &      day                               & \cite{zhu2001} \\
$k_T$              &  $10^6$                 & $\mbox{cells }(\mbox{mm}^3)^{-1}$  & (\cite{park2014}) \\
$\theta_1$         &  $10^4$                 & $\mbox{cells}(\mbox{mm}^3)^{-1}$  & estimated$\approx$1\% of $k_T$ \\
$\theta_2$         &    1                    & $\mbox{mm}^3\mbox{pg}^{-1}$  & (\cite{chiang2005}) \\
$d_T$              &  $1.27\times 10^{-2}$   &    day$^{-1}$    &  \cite{dachille2008} \\
$\delta$           &  $1.29\times 10^{-2}$   & $\mbox{mm}^3\mbox{pg}^{-1}$ & \cite{castro2004} \\
$\mu_0$            &     0.22       & $\mbox{day}^{-1}$                  & \cite{pienta1995} \\
$\mu_1$            & $9.8\times 10^{-6}$ & $\mbox{mm}^3\mbox{pg}^{-1}\mbox{day}^{-1}$ & \cite{herman2007} \\
$\mu_2$            &  $2\times 10^{-3}$  & $\mbox{mm}^3\mbox{pg}^{-1}\mbox{day}^{-1}$& \cite{magnon2013} \\
\hline
$s_G$      & $2.22\times 10^{-3}$ & $\mbox{pg}\mbox{ cell}^{-1}\mbox{day}^{-1}$  & \cite{Ldolle2003} \\
$d_G$      & 22.18                & $\mbox{day}^{-1}$                            & \cite{tang1997} \\
$\gamma_1$ & $5.57\times 10^{-5}$ & $\mbox{mm}^3\mbox{cell}^{-1}\mbox{day}^{-1}$ & \cite{rakowicz1993} \\
$\gamma_2$ & $5\times 10^{-2}$    & $\mbox{mm}^3\mbox{cell}^{-1}\mbox{day}^{-1}$ & \cite{claude1982} \\
\hline
$s_A$      & $5.06\times 10^{-3}$ & $\mbox{pg}\mbox{ cell}^{-1}\mbox{day}^{-1}$  & \cite{kigel2008} \\
$d_A$      & 2.4                  & day$^{-1}$                           & \cite{sharova2009} \\
$\gamma_3$ & $10^{-5}$& $\mbox{mm}^3\mbox{cell}^{-1}\mbox{day}^{-1}$ & estimated$\approx\gamma_4$ \\
$\gamma_4$ & $1.47\times 10^{-5}$     & $\mbox{mm}^3\mbox{cell}^{-1}\mbox{day}^{-1}$ & \cite{keino1996} \\
\hline
$r_S$            &  $6\times 10^{-2}$               & $\mbox{day}^{-1}$                  & \cite{Ldolle2003} \\
$k_S$            &        0.26        & $\mbox{cells }(\mbox{mm}^3)^{-1}$  & \cite{magnon2013} \\
$\sigma_1$       &  $1.29\times 10^2$               & $\mbox{pg day }(\mbox{mm}^3)^{-1}$ & \cite{collins1983,ruit1990} \\
$\sigma_2$       &         50                       & day                                & \cite{collins1983,ruit1990} \\
$\sigma_3$       &        7.79        & $\mbox{pg day }(\mbox{mm}^3)^{-1}$ & \cite{kuzirian2013} \\
$\sigma_4$       &        0.01        & day                           & (\cite{kuzirian2013})\\
\hline
$r_P$      &                7                 &  $\mbox{day}^{-1}$                           & \cite{collins1982,collins1983} \\
$k_P$      &  0.03   & $\mbox{cells }(\mbox{mm}^3)^{-1}$  & \cite{magnon2013} \\
$\pi_1$    &   0.33  &  $\mbox{pg}\mbox{ cell}^{-1}\mbox{day}^{-1}$ & \cite{collins1983} \\
$\pi_2$    &   0.1   &       day                                    & (\cite{collins1983}) \\
$\pi_3$    &   1     & $\mbox{pg day }(\mbox{mm}^3)^{-1}$ & estimated$\approx\sigma_3$ \\
$\pi_4$    &  0.01   & day                                & estimated$\approx\sigma_4$ \\
\hline
$c_n$      & $5\times 10^2$    & $\mbox{pg }(\mbox{mm}^3)^{-1}\mbox{day}^{-1}$ & (\cite{taubin1972}) \\
$s_n$      &      1.6          &   $\mbox{pg cells}^{-1} \mbox{day}^{-1}$       & \cite{esler1979} \\
$d_n$      &      1.66         &   $\mbox{day}^{-1}$                            & \cite{taubin1972}  \\
$\gamma_5$ & $2\times 10^{-3}$ &   $\mbox{mm}^3\mbox{cell}^{-1}\mbox{day}^{-1}$ & \cite{jaques1987} \\
\hline
$c_a$      & $2.65\times 10^4$ & $\mbox{pg }(\mbox{mm}^3)^{-1}\mbox{day}^{-1}$ & (\cite{nagler1968}) \\
$s_a$      &  0.73             & $\mbox{day}^{-1}$ & \cite{paton1971} \\
$d_a$      & 49.91             & $\mbox{day}^{-1}$ & \cite{bechem1981} \\
$\gamma_6$ & $10^{-3}$         & $\mbox{mm}^3\mbox{cell}^{-1}\mbox{day}^{-1}$  & estimated$\approx\gamma_5$ \\
\hline
\end{tabular}}

\end{small}

\caption{ A list of all the parameters appearing in the model equations.}

\label{table:paramNeuroODEs}

\end{table}

\subsubsection*{Initial conditions}
In order to explore model predictions in different scenarios we will run simulations under different initial conditions on the primary tumor cells. In particular, $T_0^{20\%}$ and $T_0^{10\%}$ denote an initial density of primary tumor cells corresponding to 20\% and 10\% of the prostate volume, respectively (see Appendix \ref{app-NeuroInitialValues} for details). 
A relatively high percentage is used due to the fact that data concerning the tumor-nerve system evolution are only available for late stages of tumor progression (as in \cite{ayala2008}).
We assume that the initial amount of tumor cells have been implanted in healthy individuals (as done in \cite{magnon2013}, although there human tumor cells were implanted in mice).
We also assume zero initial conditions for $T_m$, NGF and AGMs, because we are interested in the growth factors produced by the tumor (see above section).
All the other values are assumed to be at their normal (tumor-free) level when the model simulation starts. 

The initial values are listed in Table \ref{table:ICs-NeuroODEs}.

\begin{table}[ht]

\begin{small}

\makebox[\textwidth][c]{\begin{tabular}{cccc}
\hline
\textsc{init.value}  &  \textsc{value}   &  \textsc{units}  &  \textsc{source}   \\
\hline
     $T_p(0)$        & $T_0^{20\%}$,$T_0^{10\%}$  & cells/mm$^3$     & calculated \\
     $T_m(0)$        &       0           & cells/mm$^3$     & assumption \\
     $G(0)$          &       0           & pg/mm$^3$        & assumption \\ 
     $A(0)$          &       0           & pg/mm$^3$        & assumption \\ 
     $S(0)$          &      0.26         & cells/mm$^3$     & \cite{magnon2013} \\
     $P(0)$          &      0.03         & cells/mm$^3$     & \cite{magnon2013} \\
     $N_n(0)$        &  $3\times 10^2$   & pg/mm$^3$        & \cite{taubin1972} \\
     $N_a(0)$        & $5.3\times 10^2$  & pg/mm$^3$        & \cite{nagler1968} \\
\hline
\end{tabular}}

\end{small}

\caption{ Values of the model variables at $t=0$. }

\label{table:ICs-NeuroODEs}

\end{table}


\section{Results} 

\begin{figure}[ht]
\centerline{\includegraphics[width=0.9\textwidth]{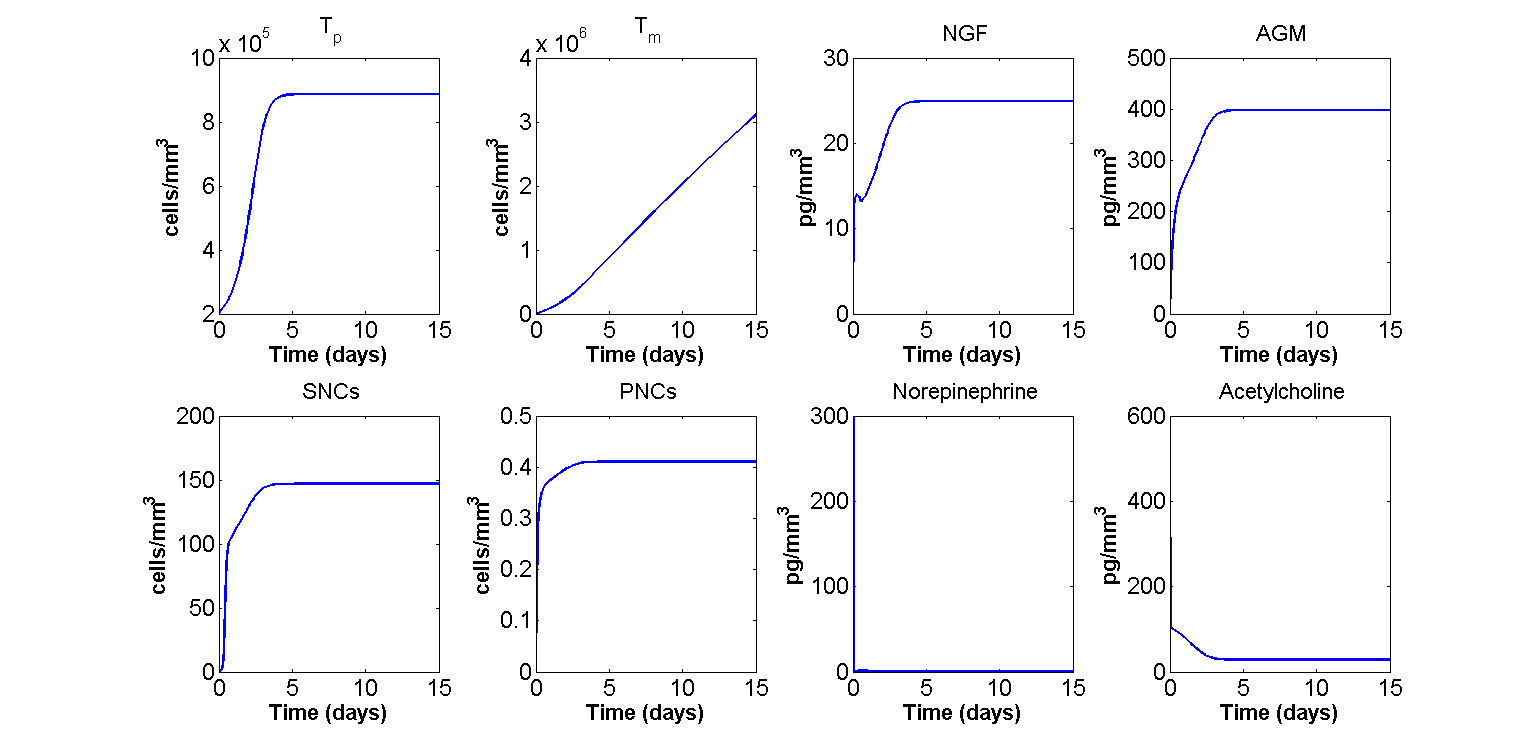}}
\caption{Time-course of the model variables over a period of 15 days for $T_0^{20\%}$.}
\label{fig:sim1}
\end{figure}

A simulation of the system of equations \eqref{eq:Tp}--\eqref{eq:Acetylch} with initial primary tumor cell density 
$T_0^{20\%}$ (see above) is shown in Figure \ref{fig:sim1}, where the MatLab function \texttt{ode45} was used 
to obtain the approximate solutions. Overall, the output is in good \emph{qualitative} agreement with the experimental observations 
associated with aggressive human prostate tumor as reported in \cite{ayala2008}. Both sympathetic and parasympathetic nerves are, in the presence of tumor, significantly increased in the region around the prostate, and the number of tumor cells leaving the domain are 
constantly increasing, matching the metastases-formation report in \cite{ayala2008,magnon2013}. Concerning the 
primary tumor mass, our model predicts that after an initial increase it reaches a nonzero equilibrium; 
this is in agreement with the results of \cite{magnon2013}, which reports an increase in tumor mass within 
the prostate. Also the fact that NGF and AGM levels stay high seems realistic: NGF levels 
are higher in inflammation \cite{watanabe2010} and some studies report that semaphorin 7A and netrin-1 
levels are significantly elevated in patients subject to chemotherapy and some kinds of 
cancers, respectively \cite{jaimes2012,ramesh2011}.
Neurotransmitters reduce rapidly to a low nonzero level following the sudden implantation of tumor cells.
On the other hand, our results are not in \emph{quantitative} agreement with \cite{magnon2013};
in particular, the present model reaches an equilibrium approximately 5 days after tumor cells implantation, whilst in \cite{magnon2013} it takes weeks to observe such significant changes. 
This may be due to the fact that the model does not take into account other elements of the prostate environment (such as lymphatic and blood vasculature) which compete with the nervous system for growth factors and space, thus potentially slowing down the dynamics.
This observation is relevant, since it implies that more elements surrounding the tumor site need to be considered in order to model tumor development more precisely.

\begin{figure}
\centerline{\includegraphics[width=0.4\textwidth]{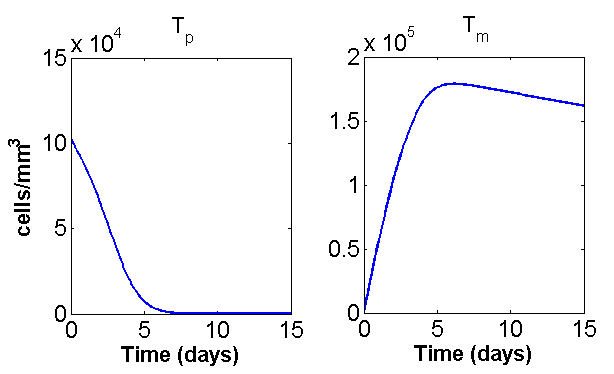}}
\caption{Primary and migrating tumor cells density time-course for initial condition $T_0^{10\%}$.}
\label{fig:sim1b-p10}
\end{figure}

An interesting feature of the model is that a smaller initial condition for primary tumor cells, for instance $T_0^{10\%}$, gives rise to completely different dynamics. In this case the primary tumor goes to zero after few days, while migrating tumor cells initially increase but then decrease to zero (Figure \ref{fig:sim1b-p10}). This behavior is in accordance with the hypothesis that a tumor cell colony has to be bigger than a certain threshold in order to proliferate \cite{korolev2014}. Note that the migrated tumor cells could cause tumor development in another site of the body where the conditions are more favorable.
It is notable that the model is able to reflect this strong dependence of tumor progression on its initial conditions; this appears to be an important feature in modern cancer research inspired by ecological dynamics \cite{korolev2014}. Our Allee threshold, lying between $T_0^{10\%}$ and $T_0^{20\%}$, appears to be unrealistically high but, to the authors' knowledge, no measurement of this parameter is available for comparison. In this model tumor cell survival and growth are affected only by nerves, while in reality blood vessels also contribute to tumor maintenance by providing oxygen and nutrients; this may (partially) account for the high threshold.

\subsection{PSA}
\begin{figure}[ht]
\centerline{\includegraphics[width=0.5\textwidth]{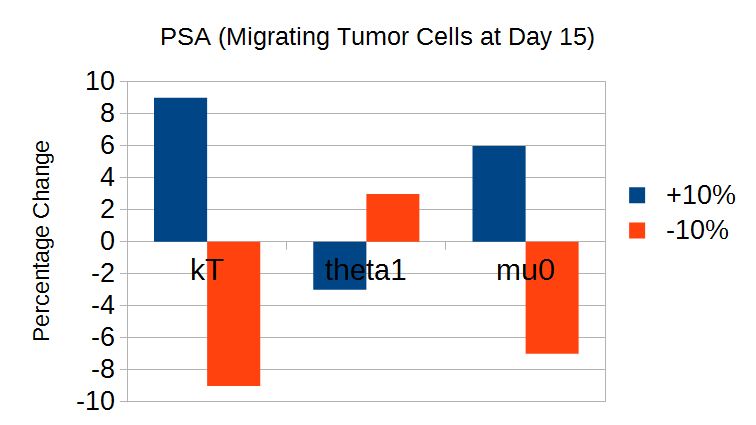}}
\caption{Parameter sensitivity analysis. The graph shows the effects on migrating tumor cells at day 15 after an increase (\textbf{{\color{blue}blue}}) or decrease (\textbf{{\color{orange}orange}}) of 10\% in the parameters. Here only the parameters which induced a percentage change of 2\% or more are shown.}
\label{fig:PSA}
\end{figure}

To test the robustness of the model, we performed a parameter sensitivity analysis (PSA) by observing the effect that a 10\% increase/reduction of each parameter value has on tumor cell densities at day 15. The model appears to be very solid in the sense that final tumor cell densities are not greatly affected by pertrbations in the parameter values. The only parameters that generate a change in migrating tumor cells of 2\% or more are reported in Figure \ref{fig:PSA}. Of these, only $k_T$ has a similar effect on primary tumor cells.

\subsection{Stress and tumor progression}
\begin{figure}[ht]
\centerline{\framebox{\includegraphics[width=0.4\textwidth]{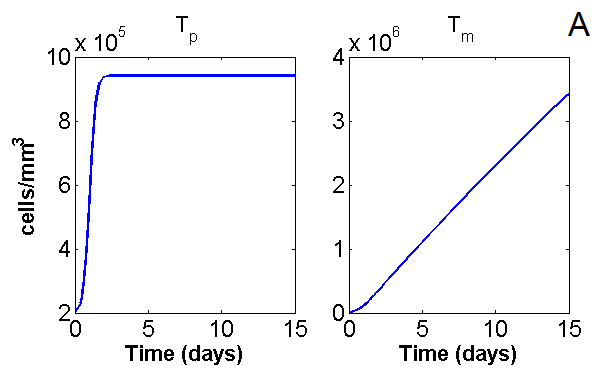}}  
 \framebox{\includegraphics[width=0.4\textwidth]{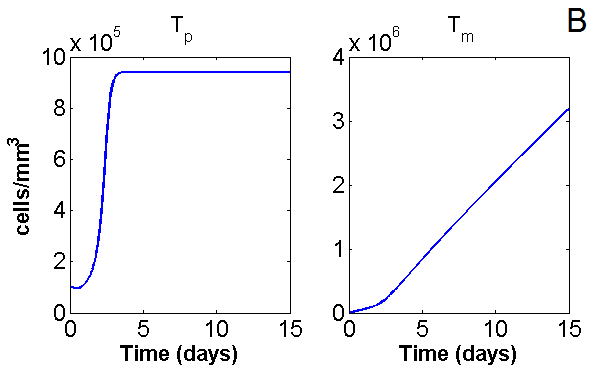}} }
\caption{Primary and migrating tumor cells in stress conditions (simulated by multiplying $s_n$ by $10$) 
         for initial conditions (A) $T_0^{20\%}$ and (B) $T_0^{10\%}$ respectively.}
\label{fig:sim2-stress}
\end{figure}

Many cancer patients exhibit stress and depression, which are known to have an effect on the immune system and consequently tumor growth \cite{mravec2008,vissoci2004}.
Additionally, they may have a direct effect as stress is associated with increased release of norepinephrine by the hypothalamus and sympathetic nerves \cite{goodmanBOOK,kvetnansky1995,pacak1995}. Here we simulate a stress condition by increasing the norepinephrine release rate $s_n$ by sympathetic nerves. Figure \ref{fig:sim2-stress}A shows the time course of primary and migrating tumor cells when $s_n$ is multiplied by 10 for initial condition $T_0^{20\%}$. The plots show that when $s_n$ is increased, the primary tumor cell density settles quickly to a higher equilibrium, while tumor cell migration is enhanced. This is in accordance with the experimental observation that stress is related to higher cancer metastasis and perhaps higher mortality \cite{chida2008,moreno2010}.
Again, our results agree qualitatively (but not quantitatively) with the experimental evidence.

Another interesting prediction of our model is that for some initial conditions, such as $T_0^{10\%}$, stress makes a crucial difference in tumor development. Here, if $s_n$ is taken to be its baseline value, recall the primary tumor tends to zero (Figure \ref{fig:sim1b-p10}); in stress conditions (simulated by multiplying $s_n$ by 10) the same initial condition leads to primary tumor growth and a constant increase of migrating tumor cells (Figure \ref{fig:sim2-stress}B).
This observation suggests that a stressful environment can affect tumor development and therapeutic efficacy, in accordance with many findings in the biological literature \cite{eng2014,vissoci2004}. More experimental data are needed to precisely quantify this effect, however this already supports the potential for treatments targeting the sympathetic nervous system, as discussed in \cite{cole2012}.

\subsection{Blocking tumor acetylcholine receptors}
\begin{figure}[ht]
\centerline{\includegraphics[width=0.4\textwidth]{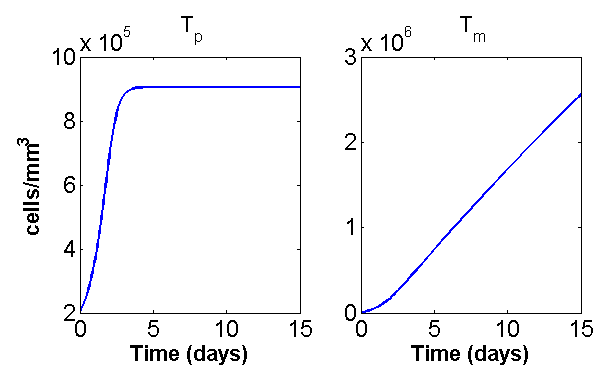}}
\caption{Primary and migrating tumor cells when acetylcholine receptors on tumor cells are blocked ($p_0=0.4$); this scenario is simulated by setting $\mu_2=0$.}
\label{fig:sim3-blockACh}
\end{figure}

Regarding parasympathetic neural activity, it is reported in \cite{magnon2013} that impairing the cholinergic (acetylcholine) receptors on tumor cells does not significantly affect tumor growth in the orthotopic site, but markedly reduces tumor cell spreading and metastasis.
To simulate this phenomenon, we set $\mu_2=0$; that is, we consider tumor cells to be non-responsive to acetylcholine. The output is shown in Figure \ref{fig:sim3-blockACh}, where we see a slower increase in the migrating tumor cells. The model corroborates the findings of \cite{magnon2013} that cholinergic receptors on tumor cells are potential clinical targets in view of limiting cancer metastasis.

\section{Discussion}  

This work is the first mathematical confirmation of the major role played by the autonomic nervous sytem in promoting tumor development and progression of prostate cancer and highlights neoneurogenesis as a target for cancer drug development. 
In the present paper we develop a simple mathematical model for tumor neoneurogenesis and cancer progression based on recent experimental evidence; it results that, regardless of the presence of angiogenesis and lymphangiogensis, tumor-induced neoneurogenesis represents a symbiotic factor for prostate tumor. This work further expands our understanding of the process by which stress can regulate cancer aetiopathogenesis: previous research predominantly emphasized the role of the immune system in mediating stress effects on tumor growth and metastasis, while our model predicts that stress can directly affect primary tumor growth through the release of neurotransmitters. 
In addition, the effect of parasympatetic nerves is also captured by the model through the acetylcholine-induced tumor migration.

This model, though quite simple, gives good insights into tumor neoneurogenesis and offers many possibilities for expansion and improvement. First of all, the introduction of a spatial variable and thus the use of PDEs would allow a more precise description of the processes occurring during tumor neoneurogenesis. In particular, a spatial approach may be able to explain why sympathetic nerves tend to accumulate in normal tissues and only penetrate tumor edges, while parasympathetic nerves infiltrate tumor tissues \cite{magnon2013}. Also, a more accurate description of the spatial component could allow for a distinction between axon elongation and nerve cell proliferation \cite{ayala2008}.

The model could be further improved by considering different variables for different kinds of AGMs, which are known to have diverse effects on tumor cells \cite{chedotal2005}. In fact, circulating tumor cells are probably attracted to a specific organ by chemokines and AGMs; the fate of a new tumor cell cluster will depend on the sensitivity of the tumor cells to the specific factors and AGMs produced in the new environment.

One could also take into account the blood and lymphatic vasculatures. 
Guidance cues for axons also have a function in (lymph)angiogenesis \cite{carmeliet2005,eichmann2005,xu2010}. Both angio-, lymphangio- and neoneuro-genesis promote metastasis formation, although in different ways; for instance, blood and lymphatic vessels offer pathways for tumor cells to disseminate, similar to perineural invasion \cite{entschladen2007}.

Another factor that could be included in the model is the immune system, which functions as a bridge between the tumor and nervous system and is the main cause of the \emph{indirect} connections between the two (in addition, NGF also seems to be involved in immune response and inflammation \cite{ehrhard1993,woolf1994}).

\section*{Acknowledgment}
The authors would like to thank Dr. Lasse Jensen, Dr. Kevin Painter and Prof. Michael Pepper for the useful technical discussions and suggestions.



\appendix
 
\section{Parameter estimation}\label{appNeuroPAR}

\subsection{Standard sizes and weights}  \label{app-NeuroSizeWeight}

\subsubsection*{Domain and normal prostate sizes}

We take normal prostate size to be approximately $30 \mbox{mL} = 3\times 10^4 \mbox{mm}^3=V_{prost}$ \cite{nickel2003}.
Assuming a spherical shape, this implies a radius of about 20 mm.

For our model, we consider the prostate and its surroundings. Therefore we consider a slightly bigger sphere, with the same centre; say (for instance) of radius 25 mm. This leads to a domain volume $V_{dom}=65.45\times 10^3 \mbox{mm}^3$.

\subsubsection*{Tumor cell size}
In \cite{park2014} the circulating tumor cells and the cultured tumor cells in prostate cancer patiens are measured; the first ones are found to have an average diameter of 7.97 $\mu$m, while the latters of 13.38 $\mu$m.
We then take a tumor cell diameter of $10\mu\mbox{m}=10^{-2}\mbox{mm}$ and thus of approximate volume $V_{tum\,cell}=5\times 10^{-7}\mbox{mm}^3$.

\subsubsection*{Neurite diameter and nerve cell size}
Take neurite diameter to be about 1 $\mu$m \cite[Table 2.1]{fiala1999}.
In \cite{friede1963} human Purkinje cell\footnote{\textbf{Purkinje cells:} a class of nerve cells} diameter is reported to be 27 $\mu$m.
We then estimate the nerve cell volume to be approximately $10^{-5}\mbox{ mm}^3$.

\subsubsection*{NGF molecular weight}
In \cite{poduslo1996} and PhosphoSitePlus NGF molecular weight is stated to be $26\times 10^3 \mbox{ Da}$.
However, NGF molecular weight is estimated to be between $10^4$ and $10^5$ Da in \cite{baker1975,murphy1977}.
We will then assume the intermediate value $10^4 \mbox{ Da}\approx 1.660\times 10^{-8} \mbox{ pg}$.

\subsubsection*{AGM molecular weight}
Molecular weight of Semaphorin 4D is 96,150 Da (see product at \texttt{www.abcam.com}).
Netrin-1 molecular weight is 75 kDa = $1.245\times 10^{-7}$ pg \cite{manitt2001}.

\subsubsection*{Noradrenaline molecular weight}
NE molecular weight = 169.17784 g/mol (from PubChem).

\subsection{Equilibrium values}  \label{app-NeuroInitialValues}

\subsubsection*{Initial primary tumor cell densities $T_0^{20\%}$, $T_0^{10\%}$}
If we take an initial density of primary umor cells corresponding to a certain percentage $p_0$ of the prostate volume, we have that
$$
T_p(0) = \frac{p_0\times V_{prost}}{V_{tum\,cell}}\times \frac{1}{V_{dom}} = p_0 \times 10^6 \; .
$$
Taking $p_0=0.20$ and $p_0=0.10$ we have the initial conditions $T_0^{20\%}$ and $T_0^{10\%}$, respectively.

\subsubsection*{Sympathetic and parasympathetic nerve density}

In \cite[Figure 7 A,D]{magnon2013} we find a quantification of sympathetic and parasympathetic (respectively) neural areas in normal human prostate tissues. From the graph, one can take a positive nerve area per field of about $1000 \mu\mbox{m}^2$ for sympathetic and $100 \mu\mbox{m}^2$ for parasympathetic fibers, field surface = $0.15 \mbox{mm}^2$. It follows that the percentage of the area occupied by nerve fibers is approximately 0.7\% and 0.07 \% for sympathetic and parasympathetic nerves respectively.
Note that here a section is 5 $\mu$m thick. However, the staining here identify any kind of nerve fibers, and it is well known that axon size is extremely variable depending on the type (for instance, in \cite{friede1963} it is recorded a nerve diameter of 27 $\mu$m, while \cite{schuman1996} reports a nerve fiber layer thickness in the eye of about 100 $\mu$m).
We will assume that the nerve fibers occupy the whole thickness of the sections; thus we conclude that sympathetic nerves account for 0.7\% of the normal prostate tissue volume and parasympathetic ones for 0.07\%.

To convert these values in an actual cells/mm$^3$ value, we recall that in \ref{app-NeuroSizeWeight} we found a domain volume of 65,450 mm$^3$. Taking the above found percentages of volume occupied by neural fibers, we have 458.1500 mm$^3$ occupied by sympathetic nerves and 45.8150 mm$^3$ by parasympathetic ones.
Approximating a nerve cell a sphere of 27 $\mu$m = $27\times 10^{-3}$ mm diameter \cite{friede1963}, we have that 458.1500 mm$^3$ correspond to 16,969 cells and 45.8150 mm$^3$ to 1,697 cells.
Therefore, the initial sympathetic nerve density will be $S^{eq}=16,969/65,450 \approx 0.26$ cells/mm$^3$ and the initial parasympathetic nerve density $P^{eq}=1,697/65,450 \approx 0.03$ cells/mm$^3$.

\subsubsection*{Noradrenaline level}
In \cite{taubin1972} the endogenous noradrenaline level in different rat gastrointestinal tissues was measured.
Taking the approximate average value 0.3 $\mu$g/g and density of the tissue equal to water's (i.e. 1g=1mL), e have that this corresponds to $N_n^{eq}=0.3\times 10^3\mbox{ pg/mm}^3$.

\subsubsection*{Acetylcholine level}
In \cite{nagler1968} small intestine and kidney acetylcholine levels are assayed in choline-deficient and normal weanling Fischer germfree\footnote{\textsc{Note:} The term ``germfree'' here refers to rats free from viable bacteria, parasites, or fungi.} and open-animal-room rats; it results that acetylcholine levels of the choline-supplemented germfree and open-animal-room rats were similar in both sites. In particular, in \cite[Table 1]{nagler1968} we find that in germfree rats acetylcholine levels in small intestine and kidneys are 0.82 $\mu$g/g and 0.24 $\mu$g/g, respectively. Taking the average of these values and considering a density equal to water's one (1g/mL) we get the ``basal'' acetylcholine level $N_a^{eq}=530$~pg/mm$^3$.


\subsection{$T_p$-equation}  \label{app-NeuroTeqn}

\subsubsection*{Tumor constant growth rate $r_T$}

In \cite{schmid1993}is reported that prostate cancer has a very large doubling time.
In particular: ``Seventy-nine percent of all patients had a doubling time of more than 24 months.
Twenty of 28 cancers thought to be clinically organ confined doubled at rates exceeding 48 months''.
We could then take $r_T = \ln 2 / 48 \mbox{months} \approx 4.81 \times 10^{-4} \mbox{day}^{-1}$.

\subsubsection*{Tumor constant death/apoptotic rate $d_T$}

In \cite{dachille2008} the apoptotic index (AI) of prostatic adenocarcinoma was calculated as 
$$
\mbox{AI (\%)} = 100 \times \mbox{apoptotic cells/total cells} \; .
$$
The mean AI in 3,000 tumor nuclei was 1.27. 
We will therefore take $d_T = 1.27\times 10^{-2}$.

\vspace{0.3cm}

\noindent
To compare these \emph{growth and death rates} with others, we see that in \cite{stein2008} it is stated that 
``The growth rate constants varied over a nearly 1,500-fold range, 
while the regression rate constants varied over a 50-fold range (Figure 3A). 
Furthermore, the regression rate constants were consistently larger than the growth rate constants, 
with median values of $10^{-1.7}\mbox{day}^{-1}$ versus $10^{-2.5}\mbox{day}^{-1}$, respectively.''
These observations correspond to $r_T\approx 10^{-2.5}\mbox{day}^{-1}$ and $d_T\approx 10^{-1.7}\mbox{day}^{-1}$. Now, while $d_T$ is approximately the same computed above, $r_T$ here is bigger; this difference is explained by the fact that prostate tumor is well-known for being particularly slow in growth.

\subsubsection*{NGF-enhanced tumor growth $\tau_1,\tau_2$}
In \cite[Figure 4A]{zhu2001} the authors report the dose-dependent effects of NGF on pancreatic cancer cell growth \emph{in vitro} after 48 hours. Here data are expressed as a percentage of increase or decrease of untreated controls. In particular, the data in Table \ref{tab:zhudata} are recorded.

\begin{table}[ht]
\begin{small}
\makebox[\textwidth][c]{\begin{tabular}{|c|c|}
\hline
NGF (ng/mL) & \% increase of untreated controls \\
\hline
    6.3     &           approx. 130             \\
     25     &           approx. 180             \\
    100     &           approx. 210             \\
\hline
\end{tabular}}
\end{small}
\caption{(Recall: 1 ng/mL = 1 pg/mm$^3$.) Time = 48 hours = 2 days.}
\label{tab:zhudata}
\end{table}

\noindent
We then consider the NGF-dependent growth part in the $T_p$-equation
$$
\frac{dT_p}{dt} = \left( r_T + \frac{G}{\tau_1 + \tau_2 G} - d_T \right) \cdot T_p
$$
that, assuming $G$ constant, has solution
\begin{equation}
T(t) = T_0 \exp \left[ \left( r_T + \frac{G}{\tau_1 + \tau_2 G} - d_T \right) \cdot t \right]
\end{equation}
that for $G=0$ reduces to
\begin{equation}
T(t) = T_0 \exp \left[ \left( r_T - d_T \right) \cdot t \right] \; ,
\end{equation}
which will correspond to the control case.

Now, from the data in Table \ref{tab:zhudata} we see that if, for example, $G=6.3$, then
$$
\frac{T_{G=6.3}(t=2)}{T_{G=0}(t=2)} = \exp \left[ \frac{G}{\tau_1 + \tau_2 G} \cdot t \right] = 1.3 \; .
$$
Similarly
$$
\frac{T_{G=25}(t=2)}{T_{G=0}(t=2)} = 1.8 \quad \mbox{and} \quad \frac{T_{G=100}(t=2)}{T_{G=0}(t=2)} = 2.1 \; .
$$
We thus have a system of three equations in two unknowns $\tau_1$, $\tau_2$:
\begin{equation} \label{eq:sys-tau1,2}
\left\{ \begin{array}{l}
\ln (1.3) \left[ \tau_1 + 6.3\cdot\tau_2 \right] = 2 \cdot 6.3  \\
\ln (1.8) \left[ \tau_1 + 25\cdot\tau_2 \right] = 2 \cdot 25 \\
\ln (2.1) \left[ \tau_1 + 100\cdot\tau_2 \right] = 2 \cdot 100
\end{array} \right.
\end{equation}
From the second equation in \eqref{eq:sys-tau1,2} we get $\tau_1 = 85.06 - 25\cdot \tau_2$.
Substituting this expression in the first equation gives $\tau_2 = 1.98$ and consequently $\tau_1=35.56$;
doing the same with the last (third) equation in \eqref{eq:sys-tau1,2} gives $\tau_2 = 2.46$ and $\tau_1=23.56$.
Taking the averages we get $\tau_1=30.74$ and $\tau_2 = 2.27$. 
However, these values give rise to an unrealistically big tumor cell growth; therefore, in our simulations we will take these two values to be multiplied by 10, thus having $\tau_1=307.4$ and $\tau_2 = 22.7$. This difference is justified by the fact that the dataset we used for our fit refers to pancreatic tumor cells, while in the paper we focus predominantly on prostate cancer.

\subsubsection*{Maximum tumor cell density $k_T$}
The maximum tumor cell density is given by $1\mbox{mm}^3/V_{tum\,cell}=2\times 10^{6}$; in fact, $k_T$ corresponds to the maximum number of tumor cells that can fit in every mm$^3$. 
Now, because of the presence of the stroma and other cells not implicitly included in the model, we will take half of this value $k_T=1\times 10^{6}$ cells/mm$^3$.

\subsubsection*{Shape of $\vartheta(N_n)$ and values of $\theta_1,\theta_2$}
We want the function $\vartheta=\vartheta(N_n)$ to be such that $\vartheta(0)\neq 0$ (to reflect the presence of an Allee threshold in the presence of a strong Allee effect) and that $\vartheta$ is a decreasing function of $N_n$ (in fact, our hypothesis is that norepinephrine lowers the Allee threshold, making the tumor more likely to proliferate).
We thus consider $\vartheta(N_n)={\theta_1}/{(1+\theta_2 N_n)}$, where $\theta_1$ and $\theta_2$ are two parameters to be determined.

For $\theta_2$, we consider \cite[Figure 1]{chiang2005}, where the time course of prostate tumor weight is shown in control mice and in mice treated with doxazosin, an $\alpha1$-adrenergic-antagonist ($\alpha$-blocker). In the plot, we observe that in the doxazosin-treated mice the tumor weight dropped down from about 5 g to zero, while in control mice a tumor of weight around 2 g kept growing.
Assuming that the doxazosin treatment blocked all the adrenergic receptors on tumor cells (thus corresponding to the case $N_n=0$), and that in the control experiment the norepinephrine was at its equilibrium value $N_n^{eq}$, we deduce that 
\begin{itemize}
\item when $N_n=0$ (i.e. norepinephrine does not make any effect on tumor growth), 5 g is \emph{below} the Allee threshold; 
\item when $N=N_n^{eq}$, 2 g is \emph{above} the Allee threshold.
\end{itemize}
Now, since it is difficult to translate these tumor weights in tumor cell densities (mouse prostate size and tumor cell size are probably different from human ones), we can only use the ``relative'' information contained above, that is 
$$
\left. \begin{array}{c}
\theta_1 > 5\mbox{ g} \\
\frac{\theta_1}{1+\theta_2 N_n^{eq}} < 2\mbox{ g} \end{array} \right\} \: \Rightarrow \: 1 + \theta_2 N_n^{eq} > \frac{5}{2} \: \Rightarrow \: \theta_2 > 5\times 10^{-3} \frac{\mbox{mm}^3}{\mbox{pg}} \; .
$$
We can take for instance $\theta_2=1\mbox{ mm}^3/\mbox{pg}$.

As pointed out in \cite{korolev2014}, no experiment has been done to measure the ``basal'' Allee threshold $\theta_1$ for any kind of tumor. We will just assume that $\theta_1$ is approximately the 1\% of the carrying capacity $k_T$, i.e. $\theta_1=1\times 10^4$.

\subsubsection*{AGM-induced tumor cell apoptosis $\delta$}
In \cite[Table 1]{castro2004} we find a quantification of the effect of semaphorin 3B on two different kinds of cancer cells; these data are summarised in Table \ref{tab:castrodata}.

\begin{table}[ht]
\begin{small}
\makebox[\textwidth][c]{\begin{tabular}{|c|c|c|}
\hline
Treatment  & H1299 lung cancer cells & MDA-MB-231 breast cancer cells \\
\hline
Control-CM &     $11\times 10^4$     &       $16\times 10^4$   \\
SEMA3B-CM  &      $6\times 10^4$     &        $5\times 10^4$   \\
\hline
\end{tabular}}
\end{small}
\caption{Time = 5 days; $C_0 = 10^4$ cells/well (six-well plates)}
\label{tab:castrodata}
\end{table}

\noindent
We will then consider the following two equations for control tumor cells $T_{control}$ and for semaphorin-treated ones $T_{SEMA}$:
$$
T_{control}(t) = T_0 \exp \left[ \left( r_T - d_T\right) t\right] \; \mbox{ and } 
\; T_{SEMA}(t) = T_0 \exp \left[ \left( r_T - d_T - \delta A\right) t\right] \; ,
$$
where $A$ represents the concentration of axon guidance molecule (here, semaphorin).
To estimate $A$ we consider the statement ``Semiquantitative assay showed an average of 15–40 ng/mL SEMA3B in theCMafter transfection'' in the \emph{Materials and Methods} section and the fact that the medium was diluted 1:2; in this way we approximate $A\approx 13.75\mbox{ pg/mm}^3$ (note that 1 ng/mL = 1 pg/mm$^3$).
Equipped with all these values (recall: $t=5$), we can use the data in Table \ref{tab:castrodata} as follows:
$$
\mbox{H1299 cells: } \frac{T_{SEMA}}{T_{control}} = \exp \left( -\delta A t\right) = \frac{6}{11} \, ,
\, \mbox{MDA-MB-231 cells: } \frac{T_{SEMA}}{T_{control}} = \frac{5}{16}
$$
and then calculate the corresponding $\delta$ values 0.0088 and 0.0169 respectively. Taking the average, we get $\delta \approx 1.29\times 10^{-2}$.

\subsubsection*{Spontaneous tumor cell migration $\mu_0$}

In \cite{pienta1995} about 1,400 colonies of (rat) prostate tumor cells are observed after 8 days (see Figure 4 in the same reference).
Without knowing how big each colony is, we will assume that 1 colony corresponds to 1 cell. Therefore, taking an exponential decay $T_p(t)=T_p(0)\exp\left(-\mu_0 t\right)$ for the tumor cells and knowing that the initial cell density was $T_p(0)=2\times 4\times 10^3$ cells/mL (stated also in \cite{pienta1995}), we can calculate $\mu_0 = 0.22\mbox{ day}^{-1}$.

\subsubsection*{AGM-induced migration $\mu_1$}
In \cite[Figure 3]{herman2007} the following \% cell invasion are reported for semaphorin-treated PC-3 cells\footnote{\textbf{PC-3 cells:} androgen-independent prostate cell line.}:
$$
\mbox{sema3A: } \sim 65\% \mbox{ of control } \quad , \quad \mbox{sema3C: } \sim 135\% \mbox{ of control } 
$$
after 20 hours incubation ($T_0=10^5$).
The authors' comment is: ``Overexpression of sema3A in PC-3 decreased the invasive characteristics of PC-3 cells by 33\% compared to the untransfected cells. Sema3C, on the other hand, increased invasion by 33\% compared to untransfected cells''.
To estimate the amount of semaphorin used in the experiment, we read: ``The bacterial clones transfected with sema3A or sema3C were grown on agar plates and selected with 35 $\mu$g/mL of kanamycin''. Therefore, in our equation for tumor cell migration $T_m(t)=T_0\exp[(\mu_0+\mu_1 A) t]$ we will take $A=35\,\mu\mbox{g/mL}=35\times 10^3\mbox{ pg/mm}^3$.
Finally, considering the 20-hours sema3C treatment, we have that
$$
\frac{T_0\exp\left[(\mu_0+\mu_1 A) t\right]}{T_0\exp\left(\mu_0 t\right)} = 1.33 \quad \Rightarrow
\quad \mu_1 A t = \ln (1.33) \quad \Rightarrow \quad \mu_1 = 9.8\times 10^{-6} \frac{\mbox{mm}^3}{\mbox{pg}\cdot\mbox{day}} \; .
$$

\subsubsection*{Acetylcholine-induced migration $\mu_2$}
\cite[Figure 3A]{magnon2013} reports an \emph{ex vivo} quantification of tumor cell invasion of pelvic limph nodes (which drain the prostate gland). Here data are reported both for control (saline-treated) and carbachol-treated mice, and in the second case the invading tumor cells are about the double than in the control. Notice that since carbachol is a non-selective cholinergic (muscarinic) receptor agonist, we can consider it as a substitute of acetylcholine. Then, denoting with $c$ the carbachol amount, we can estimate from the equation $T_m(t)=T_0\exp[(\mu_0+\mu_2 N_a) t]$ and \cite[Figure 3A]{magnon2013} that $\mu_2 c t = \ln (2)$ (since $T_0\exp[(\mu_0+\mu_2 c) t]\approx T_0\exp(\mu_0 t)$). To estimate the value of $c$, we read in \cite{magnon2013}: ``For  experiments on the PNS, 15 days after tumor cell injection, animals received carbachol at 250 (day 0), 300 (day 1), 350 (day 2), 500 $\mu$g/kg per day (day 3) [every 12 hours, 8 divided doses]''.
First notice that the average of these amounts is 350 $\mu$/kg over 5 weeks, which corresponds to 10 $\mu$g/kg/day.
To convert the kilos in a volume, we take water density (1 g/mL); therefore we find the approximation $c=10\mbox{ pg/mm}^3\mbox{/day}$ and thus $\mu_2 = 2\times 10^{-3}\mbox{ mm}^3\mbox{pg}^{-1}\mbox{day}^{-1}$ ($t=35$ days).

\subsection{$G$-equation}  \label{app-NGF}

\subsubsection*{NGF decay rate $d_G$}
In \cite{tang1997} is stated that ``Nerve growth factor (NGF) mRNA is rapidly degraded in many non-neuronal cell types with a half-life of between 30 and 60 min''.
Hence, taking a half-life of 45 minutes, 
the resulting decay rate is $d_G = 0.0154 \mbox{ min}^{-1} = 22.18 \mbox{ day}^{-1}$.

\subsubsection*{NGF production by tumor cells $s_G$}
In \cite[Figure 1c]{Ldolle2003} it is reported that after 24 hours, cultures of different lines of breast cancer cells expressed approximately 0.3 ng/(mg protein) of NGF. Considering a total protein amount of 300 pg per cell (as in 
HeLa cells\footnote{\textbf{HeLa cells:} an immortalised cell type used in biological research, derived from cervical cancer cells taken from Henrietta Lacks, who eventually died of her cancer in 1951.}),
we have that 1 mg = $10^9$ pg protein corresponds to approximately $3\times 10^6$ cells.
Now, we have to consider that in 24 hours the NGF also decayed; in fact the differential equation for $G$ in this case is 
$$
\frac{dG}{dt} = s_G T - d_G G \quad \Rightarrow \quad G(t) = \left( G(0) - \frac{s_G}{d_G}T \right) \exp (-d_G t) + \frac{s_G}{d_G}T
$$
where $T$ denotes the number of tumor cells (and $G(0)=0$ in our case).
Thus, \cite{Ldolle2003} tells us that $G(t=1\mbox{day})=0.3\times 10^3\mbox{pg}$, $T=3\times 10^6\mbox{cells}$.
Substituting these numbers in the equation (and taking the value of $d_G$ estimated above), we determine $s_G = 2.22\times 10^{-3} \mbox{ pg}\cdot\mbox{cells}^{-1}\cdot\mbox{day}^{-1}$. 


\subsubsection*{NGF internalization rate by tumor cells $\gamma_1$}
In \cite[Table 1]{rakowicz1993} it is reported the internalization of \textsuperscript{125}I-NGF after 1 hour or 24 hours incubation of different cell lines with 10 ng/mL. For SKBr5 breast carcinoma cells, we find that 33,560 molecules/cell were internalised after 24 hours incubation.
Considering a NGF molecular weight of $1.660 \times 10^{-8}\mbox{ pg}$ and knowing that the cells were seeded at density $2\times 10^7 \mbox{ cells/10 mL} = 2\times 10^3 \mbox{ cells/mm}^3$, we can wite down the equality
$$
\gamma_1 \times \left( 2\times 10^3 \frac{\mbox{cells}}{\mbox{mm}^3} \right)\times \left( 10 \frac{\mbox{pg}}{\mbox{mm}^3} \right)
 = 33.56 \times 10^3 \times 1.660 \times 10^{-8} \frac{\mbox{pg}}{\mbox{mm}^3} \times 2 \times 10^3 \mbox{ cells} \frac{1}{\mbox{day}}  \; ,
$$ 
which leads to $\gamma_1 = 5.57\times 10^{-5} \mbox{ mm}^3\mbox{cells}^{-1}\mbox{day}^{-1}$.

\subsubsection*{NGF internalization rate by nerve cells $\gamma_2$}
We can estimate the rate of NGF internalization by cultured neurons using the data in \cite[Figure 1]{claude1982}.
The plot reports the pg of \textsuperscript{125}I-NGF binding to rat sympathetic neurons  versus  different amounts of free NGF.
It is also stated that the neurons were incubated for 140 minutes with the NGF at a density of approximately 1,000 neurons/dish in 35-mm culture dishes.
 
\begin{figure}[h!]
      \centering
      \includegraphics[height=7cm]{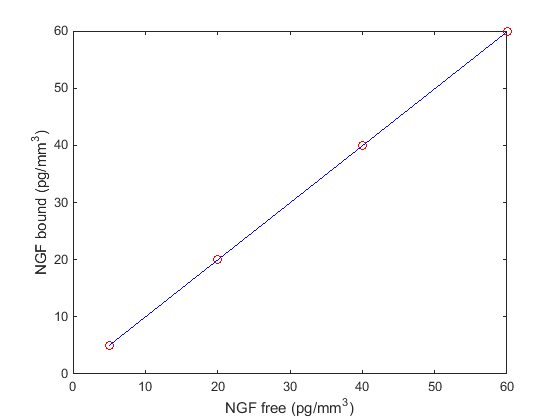}
      \caption{Plotting the data from \cite{claude1982} ({\color{red}red circles}) 
      together with the function $G(t)=G_0\exp(-\gamma_2 S t)$ ({\color{blue}blue line}) 
      fitted to the data with  the MatLab function \texttt{nlinfit}.}
       \label{fig:claudeFIT}
\end{figure}

Therefore, if we have a density of free NGF equal to $G_0$, the corresponding value on the $y$-axis of \cite[Figure 1]{claude1982} corresponds to $G(t=140\mbox{min})=G_0\exp(-\gamma_2 St)$. Then, converting these data into our units, we can use the MatLab functions \texttt{nlinfit} and \texttt{nlparci} to get an estimate for $\gamma_2$ and its 95\% confidence interval respectively.
The plot of the fit is reported in Figure \ref{fig:claudeFIT} and the output is:

\begin{verbatim}
>> claude.ngf = [5 20 40 60]
>> claude.binding = [4.98 19.95 39.90 59.865]
>> NGFinternFIT(claude)
Estimated gamma2 value = 0.048342
ci for gamma2 =    0.0422    0.0545
\end{verbatim}

\subsection{$A$-equation}  \label{app-AGM}

A large class of secreted or membrane bound axon guidance molecules are semaphorins and more specifically the so called class-3 semaphorins, that include seven family members. Class 3 semaphorins are the only secreted vertebrate semaphorins. 
In a recent work, the authors of \cite{blanc2011} highlighted that Semaphorin 3E is not only overexpressed in prostate cancer but also affects 
adhesion and motility of prostate cancer cells. They also demonstrated that all the prostate cancer cell lines that have been tested produce both the unprocessed (87kDa) and processed (61kDa) form of Sema3E. However the effect of tumor and stromal secreted semaphorins on tumor functionalities such as migration, apoptosis, growth and invasion is likely to depend on which co-receptors are expressed. Namely, sema3E act as a chemoattractant fro neurons expressing NRP1 receptors, that have been found to have a high expression on prostate tumors.

\subsubsection*{AGM tumor secretion rate $s_A$}

In \cite{kigel2008} the concentration of secreted sema3s in conditioned medium is estimated for specific tumor cell lines. As it can be deduced by \cite[Figure 2]{kigel2008} the relative concentrations of class-3 semaphorins secreted into the medium of tumor cell lines were 1000 and 500 sema3-expression per cell. Tumor cells were incubated for 48 hours = 2 days. Take an average of the aforementioned values, we deduce that the expression of sema3 pre-cell per-day is 350. Taking the molecular weight of unprocessed sema3 to be 87kDa (as described at the beginning of this section \ref{app-AGM}), we estimate that the secretion rate is: 
$s_A = 350 \times 87000 \times 1.66 \times 10^{-12} \mbox{pg cell}^{-1} \mbox{day}^{-1}$, thus 
$s_A \approx 5.055 \times 10^{-5} \mbox{pg cell}^{-1} \mbox{day}^{-1}$. 
However, in \cite{kigel2008} it is highlighted that the aforementioned expressed semaphorins did not affect the proliferation rate or the 
survival of the different semaphorin tumor producing cells. In this regard, we expect that during tumor driven neo-neurogenesis the expressed tumor secreted sema3E are 100 or 1000 higher then the estimated value, in other words we take $s_A = 5.06 \times 10^{-3} \mbox{pg cell}^{-1} \mbox{day}^{-1}$.

\subsubsection*{AGM decay rate $d_A$}
In \cite[Supplementary Tables 1 and 2]{sharova2009} we find the mRNA half-life of different kinds of semaphorins.
We take an average decay rate of $0.1 \mbox{ h}^{-1} = 2.4 \mbox{ day}^{-1}$.

In \cite{manitt2001} we read: ``Currently, little is known about the half-life of netrin-1 protein in any context''.


\subsubsection*{AGM internalization by nerve cells $\gamma_4$}
\begin{figure}[h!]
      \centering
      \includegraphics[height=7cm]{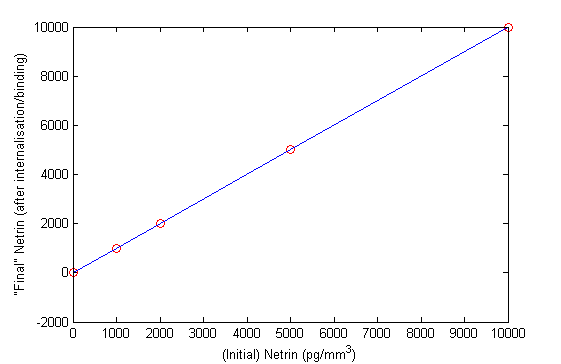}
      \caption{Plotting the data from \cite{keino1996} ({\color{red}red circles}) 
      together with the function $A(A_0)=A_0\exp(-\gamma_4 S t)$ ({\color{blue}blue line}) 
      fitted to the data with  the MatLab function \texttt{nlinfit}.}
       \label{fig:keinoFIT}
\end{figure}

In \cite[Figure 4]{keino1996} it is studied the binding of netrin(VI$\bullet$V)-Fc to DCC-expressing cells (spinal commissural axons).
Here the c.p.m. are reported for different concentrations of netrin. 
Assuming the every binding corresponds to 1 molecule, and taking the netrin molecular weight $1.245\times 10^{-7}$ pg, we can calculate the decrease of \emph{free netrin}, that in our system is represented by the variable $A$ and in this case is modelled by the equation $A(t)=A_0\exp(-\gamma_4 S t)$. Then, having $S=2.5\times 10^5$ cells/24-well = 71.43 cells/mm$^3$ (24-well $\rightarrow$ 3.5 mL) and $t = 1$ minute = $6.9444\times 10^{-4}$ days, we can fit this as a function of $A_0$, as in \cite[Figure 4]{keino1996}.
The MatLab functions \texttt{nlinfit} and \texttt{nlparci} give the following $\gamma_4$ estimate and 95\% confidence interval:
\begin{verbatim}
>> keino.netrin = [0 1 2 5 10].*10^3
>> keino.binding = [0.5 1.5 3.5 4.5 4.5].*10^4
>> AGMinternFIT(keino)
Estimated gamma4 value = 1.4673e-005
ci for gamma4 =    1.0e-004 * [    0.0462    0.2472    ]
\end{verbatim}

\noindent
The plot of this fitting is reported in Figure \ref{fig:keinoFIT}.

\subsection{$S$-equation}  \label{app-SNC}

\subsubsection*{Basal SNC growth rate $r_S$}
In \cite[Table 1]{Ldolle2003} we find that 4.4\% control sympathetic neurons cultured for 48 hours showed a neurite length of 29 mm. The initial cell density was $2\times 10^3$ cells/well that, assuming a well volume of 100 mL, correspond to $S_0=2\times 10^{-2}\mbox{ cells/mm}^3$; moreover, taking a neurite diameter of 1 $\mu$m, we have that 29 mm neurite correspond o approximately 2.9 cells (recalling that we consider a nerve cell volume of about $10^{-5}\mbox{mm}^3$).

From \cite{Ldolle2003}: ``Cell culture plates (96-well) were prepared by incubating each well with 100 mL of 0.1mg/mL poly-L-lysin in sterile distilled water [...] Approximately $2\times 10^3$ cells, prepared from embryonic day-12 chick paravertebral sympathetic ganglia, were added to each well in 100 mL of a 1:1 mixture of [...] medium''.

Hence, we conclude that after 2 days of experiment there were
$$
S(t=2) = S_0 + 0.044\times S_0 \times 2.9 \mbox{ cells} \quad \Rightarrow 
\quad S(t=2)=  2.2552 \times 10^{-2}\mbox{cells/mm}^3 = S_0 \exp\left(2\cdot r_S\right) \; ,
$$
from which we can calculate $r_S = 0.06\mbox{ day}^{-1}$.

\subsubsection*{SNC carrying capacity $k_S$}
In absence of tumor, we know that the SNC equilibrium value is $S^{eq}=0.26$ cells/mm$^3$ (see section \ref{app-NeuroInitialValues}).
We then take $k_S=S^{eq}$.

\subsubsection*{NGF-enhanced SNC growth  $\sigma_1,\sigma_2$}
In \cite[Table 1]{ruit1990} the effects of NGF treatment on superior cervical ganglion cell dendritic morphology are reported; they are summarised here in Table \ref{tab:ruitdata}.

\begin{table}[ht]
\begin{small}
\makebox[\textwidth][c]{\begin{tabular}{|ccc|}
\hline
Treatment & Animal size & Total dendritic length ($\mu$m) \\
\hline
Control   &   23.5 g    &    721            \\
NGF       &   23.5 g    &    929            \\
\hline
\end{tabular}}
\end{small}
\caption{Mouse 2.5S NGF was administered daily to mice by subcutaneous injection in a dosage of 5 mg/kg. The animals were treated for 2 weeks.}
\label{tab:ruitdata}
\end{table}

\noindent
Now, if we take a dendritic diameter of 1 $\mu$m and a nerve cell volume of $10^{-5}\mbox{ mm}^3$, we have that 1 $\mu$m dendrite corresponds to about $10^{-4}$ cells. Therefore, we can ``convert'' the previous dendritic lengths in cells (at least roughly). 
For the NGF treatment, we know that it was 5 mg/kg/day for 2 weeks. If a mouse was 23.5 g, we have that each animal received $117.50\times 10^6$ pg/day. Being NGF injected subcutaneously, we assume that only 1\% of the dosage actually contributed to the experiment (the rest being dispersed by body fluids).
Additionally, we estimate a total mouse volume of $28.57\times 10^3\mbox{ mm}^3$ (knowing that mice blood volume is about 2 mL and it constitutes 7-8\% of their total volume \cite{JHU-Mouse}) and thus we have a daily NGF supply of 41.13 pg/mm$^3$/day.
Now, to calculate the effective NGF present, we have to take into account its decay. We know that NGF decay rate is $d_G = 22.18\mbox{ day}^{-1}$ (see \ref{app-NGF}); if we define the constant supply $s = 41.13 \mbox{ pg }(\mbox{mm}^3)^{-1}\mbox{day}^{-1}$, we have that the evolution equation for $G$ in this setting is
$$
\frac{dG}{dt} = s - d_G G \quad \Longrightarrow \quad G(t) = G_0 e^{d_G t} - \frac{s}{d_G}e^{-d_G t} + \frac{s}{d_G} \; .
$$
Then, taking $G_0=0$, we have that at $t=1\mbox{ day}$ the amount of NGF is approximately $1.85 \mbox{ pg/mm}^3$. For the two-week experiment, we will then assume $G$ to be $1.85\times 14 = 25.96 \mbox{ pg/mm}^3$.
Then, back to the $S$-equation: we recall that the bit in which we are now interested is
$$
\frac{dS}{dt} = \left( r_S + \frac{G}{\sigma_1 + \sigma_2 G} \right) S \quad \stackrel{G\,\mbox{\footnotesize const}}{\Longrightarrow}
\quad S(t) = S_0 \exp \left[ \left(r_S + \frac{G}{\sigma_1 + \sigma_2 G}\right)t\right] 
$$
we have that at 2 weeks = 14 days
$$
\frac{S_{NGF}}{S_{control}} = \exp \left( \frac{G}{\sigma_1 + \sigma_2 G}\times 14 \right) = \frac{929}{721} = 1.29 \; .
$$
From this equation (recall: $G = 25.96$) we derive $\sigma_1 = 25.96\times(54.9791-\sigma_2)$.
Consequently, we have that it must be $\sigma_2 < 54.9791$ in order to have $\sigma_1 > 0$.

We can derive a second equation for $\sigma_1$ and $\sigma_2$ from the experimental results reported in \cite{collins1983}.
In fact, \cite[Table 1]{collins1983} lists the maximal effects on neurite lengths of various additions to the culture medium. In particular, the mean total neurite length per neuron after different treatments divided by the corresponding value of the untreated control is reported. For sympathetic neurons exposed for 2 hours to 1 ng/mL NGF, the relative length is 2.47; this observation allows us to write the following equality:
$$
\exp\left( \frac{G}{\sigma_1+\sigma_2 G}\times 0.0833 \right) = 2.47 \quad \Longrightarrow 
\quad \sigma_2 = 57.1782 \; ,
$$
the latter obtained after substituting the expression for $\sigma_1$ found previously (note that 2 hours = 0.0833 days).
Notice that although $\sigma_2$ is bigger than 54.9791, the difference is small (less than one order of magnitude). This is probably due to the fact that the two references used to estimate $\sigma_1$, $\sigma_2$ deal with completely different experimental settings (for example, the experiment done in \cite{ruit1990} is \emph{in vivo} while \cite{collins1983} is \emph{in vitro}). Therefore it seems justified to take for instance $\sigma_2 = 50$ days and consequently $\sigma_1 \approx 129 \mbox{ pg day }(\mbox{mm}^3)^{-1}$.


\subsubsection*{AGM-enhanced SNC growth $\sigma_3,\sigma_4$}

In \cite[Figure 1A($ii$) and 2A]{kuzirian2013} the synapse density after 0.5, 1, 2 and 4 hours of Sema4D treatment is reported as \% of control. In particular, it is reported that after 0.5 hours = 0.0208 days = $t_1$ of 1nM-Sema4D-treatment GABAergic synapse formation in rodent hippocampus was about 130\% of control, and after 1 hour = 0.0417 days = $t_2$ it was approximately 150\% of  control.
Now, recalling the ``growth bit'' of the $S$-equation
$$
\frac{dS}{dt} = \left( r_S + \frac{A}{\sigma_3 + \sigma_4 A} \right) S \quad \stackrel{A\,\mbox{\footnotesize const}}{\Longrightarrow}
\quad S(t)= S_0 \exp \left[ \left( r_S + \frac{A}{\sigma_3 + \sigma_4 A} \right) t \right] \; ,
$$
we have from the previous datapoints that
$$
\exp \left( \frac{A t_1}{\sigma_3 + \sigma_4 A} \right) = 1.3 \quad \mbox{ and } \quad \exp \left( \frac{A t_2}{\sigma_3 + \sigma_4 A} \right) = 1.5
$$
(since the control corresponds to the $S(t)$ where $A=0$).
Note that taking molecular weight of 96,150 Da for $A$, we have that $A=1\mbox{ nM}=96.117$ pg/mm$^3$.
Finally, considering the average of the two expressions above we can estimate
$$
\sigma_3 + \sigma_4 A \approx \frac{1}{2}\left( \frac{At_1}{\ln (1.3)} + \frac{At_2}{\ln (1.5)}\right) \quad \Rightarrow \quad \sigma_3 \approx 8.75 - 96.12\times \sigma_4 \; .
$$
Note that we must choose $\sigma_4 < 0.0911$ in order to have $\sigma_3 > 0$.
Taking for instance $\sigma_4 = 0.01$, we have consequently also $\sigma_3 = 7.79$.



\subsection{$P$-equation}

\subsubsection*{Basal PNC growth rate $r_P$}
In \cite[Table I]{collins1982} it is reported that the mean total neurite legth/neuron after $2\nicefrac{3}{4}$ hours in conditioned medium was 408 $\mu$m, while  in the unconditioned medium it was 118 $\mu$m (they study ciliary ganglia, which are parasympathetic ganglia located in the posterior orbit). Taking the latter as the initial value $P_0$, we have from the equation describing PNC dynamics in this context
\begin{equation*}
\begin{split}
&P(t=2\nicefrac{3}{4}\mbox{h}=0.1146\mbox{day}) = P_0 \exp \left( r_P \times t \right) \\
& \Rightarrow \quad 0.1146\times r_P = \ln \left( \frac{408}{118} \right) \\
& \Rightarrow \quad r_P = 10.83 \mbox{ day}^{-1} \; .
\end{split}
\end{equation*}

In the same reference we find another useful dataset in \cite[Table II]{collins1982}. Here it is stated that the mean elongation rate of 14 neurites (chosen to be at east 15 $\mu$m long) without any medium change was 22 $\mu$m/hour.
Converting these lengths into cell numbers (using the calculations done in \ref{app-NeuroSizeWeight}) and keeping in mind that 1 hour = 0.0417 days, we calculate the growth rate ``per cell'' $r_P$ as $\nicefrac{22}{14\times 15}\times \nicefrac{1}{0.0417}=2.51\mbox{ day}^{-1}$.

Another way to determine $r_P$ could be to use the data in \cite[Table 1]{collins1983}. Here the authors measure the maximal effect on ciliary (parasympathetic) and sympathetic neurite growth in various culture media after 2 hours. Considering the data regarding the ``standard'' conditioned medium, we have that the relative neurite length for ciliary neurons was 3.42, and for sympathetic neurons 1.81. Then, assuming an exponential growth for both cell cultures, we have that ${P_0\exp(r_Pt)}/{S_0\exp(r_St)}=3.42/1.81$; furthermore, taking $P_0 = S_0$ and $t=2\mbox{h}=0.0833\mbox{day}$, we have that $r_P-r_S=7.63\mbox{ day}^{-1}$.
Now, recalling our previous estimate for $r_S$ ($r_S=0.06$, see \ref{app-SNC}), we have $r_P=7.70\mbox{ day}^{-1}$.

It is encouraging to see that all these three values are of the same order of magnitude. To choose an estimate for $r_P$, we take their average $7\mbox{ day}^{-1}$.

\subsubsection*{PNC carrying capacity $k_P$}
In absence of tumor, we know that the PNC equilibrium value is $P^{eq}=0.026$ cells/mm$^3$ (see section \ref{app-NeuroInitialValues}).
We then take $k_P=P^{eq}$.

\subsubsection*{NGF-enhanced PNC growth $\pi_1 , \pi_2$}
\cite{collins1983} investigated (in vitro) the effect of NGF on promoting the parasympathetic ciliary ganglion outgrowth.  
Their calculations were used to calculate the mean total length of neurites per neuron. Their calculations were based on data 
from neurons that had at least one neurite greater then 15 $\mu m$ in length ($\approx$ about the diameter of the neuronal 
soma). In this regard when they added NGF to dissociate ciliary ganglion neurons, resulted in a 2-fold increase in neurite length 
over untreated, control cultures.They estimated the mean total neurite length per neuron for control cultures to be 
$79 \pm 19 \mu\mbox{m}$. Parasympathetic ganglion neurons were exposed to a concentration of 
$10 \mbox{ng/mL} = \frac{10 \times 10^{3}}{10^{3}}$ $\frac{\mbox{pg}}{\mbox{mm}^{3}}$
per h. Just two hours after addition of NGF the ratio $\frac{P_{NGF}}{P_{control}} \approx 2.08 \pm 0.12$. 

Recalling the given $P$ equation:

$$
\frac{dP}{dt} = \left( r_P + \frac{G}{\pi_1 + \pi_2 G} \right) S \quad \stackrel{G\,\mbox{\footnotesize const}}{\Longrightarrow}
\quad P(t) = P_0 \exp \left[ \left(r_P + \frac{G}{\pi_1 + \pi_2 G}\right)t\right]  \; ;
$$
so after two hours we have
$$
\frac{P_{NGF}}{P_{control}} = \exp \left( \frac{G}{\pi_1 + \pi_2 G}\times \frac{2}{24} \right) = 2.08 \; .
$$
Taking into account that 2 hours $\approx$ $\frac{2}{24} \mbox{day} = 0.083 \mbox{day}$ we deduce that
$$
 \frac{10}{\pi_1+\pi_2\times 10} \times 0.083 = \ln 2.08
$$
and therefore $\pi_1 =1.33 - 10 \times \pi_2$.
Note that it must be $\pi_2 < 0.13$ in order to have $\pi_1 > 0$.
We can take for example $\pi_2=0.1$ and thus $\pi_1=0.33$.



\subsection{$N_n$-equation}

\subsubsection*{Noradrenaline production by SNC $s_n$}
Regarding the norepinephrine release rate, \cite{esler1979} estimated the apparent norepinephrine release rate at rest to be $0.54 \pm 0.20 \mu\mbox{g/(m}^2\mbox{min)}=777.60\,\nicefrac{\mbox{pg}}{\mbox{mm}^2  \times \mbox{day}}$. Note that 90\%  of this release rate is due to the sympathetic nerves. 
To convert the mm$^2$ in cells, we assume once again a nerve cell radius $r = 13.5 \mu\mbox{m} = 13.5 \times 10^{-3} \mbox{m}$ (see \ref{app-NeuroSizeWeight}); the surface area is given by $4 \pi r^{2} = 4 \pi (13.5 \times 10^{-3})^{2} \approx 2.29 \times 10^{-3} \mbox{mm}^2$, thus we deduce that in $1\mbox{ mm}^2$ there are ${1}/{(2.29 \times 10^{-3})}=436.7$ nerve cells. 
The noradrenaline secretion rate is then given by $s_n = 0.9 \times \nicefrac{777.60}{436.7} \approx 1.6 \mbox{ pg cells}^{-1} \mbox{day}^{-1}$. 


\subsubsection*{Noradrenaline decay rate $d_n$}

%

In \cite{taubin1972} the noradrenaline half-life is reported to be about 10 hours (although this value is different in different tissues). This leads to a decay rate $d_n = 1.66\mbox{ day}^{-1}$.

\subsubsection*{Noradrenaline uptake rate (by tumor cells) $\gamma_5$}

\begin{figure}[h!]
      \centering
      \includegraphics[height=7cm]{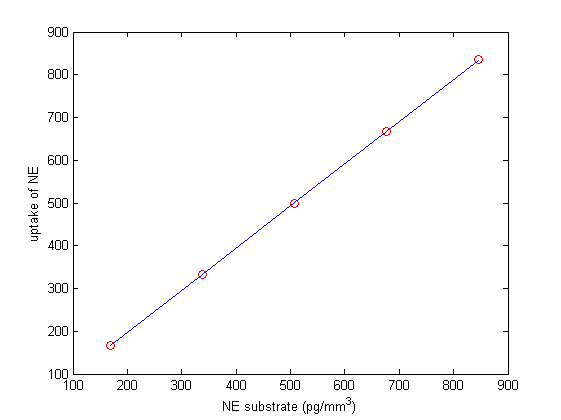}
      \caption{Plotting the data from \cite{jaques1987} ({\color{red}red circles}) 
      together with the function $N(t)=N_0-N_0\exp(-\gamma_5 T t)$ ({\color{blue}blue line}) 
      fitted to the data with  the MatLab function \texttt{nlinfit}.}
       \label{fig:jaquesFIT}
\end{figure}

In \cite[Figure 4A]{jaques1987} we find one set of measurements of NE uptake by human pheochromocytoma\footnote{A pheochromocytoma is a neuroendocrine tumor of the medulla of the adrenal glands; it secretes high amounts of catecholamines, mostly norepinephrine, plus epinephrine to a lesser extent.} cells.
Recalling the molecular weight of NE found in \ref{app-NeuroSizeWeight} and assuming a culture volume of 1 mL (it is not better specified in the paper), we can convert the datapoints in \cite[Figure 4A]{jaques1987} into our units and fit the function $N(t)=N_0-N_0\exp(-\gamma_5 T t)$ to them; note that $T$ represents the tumor cells, and that the value of this function at each time $t$ is measured as the initial substrate concentration minus the uptaken NE.
Using the MatLab function \texttt{nlinfit} to fit the data gives:

\begin{verbatim}
>> jaques.substr = 169.18.*[1 2 3 4 5]
>> jaques.uptake = 169.18*10^(-3).*[15 32 45 53 65]
>> NEinternFIT(jaques)
Estimated gamma5 value = 0.0019926
ci for gamma5 =    0.0018    0.0022
\end{verbatim}

\noindent
The plot of the fit is reported in Figure \ref{fig:jaquesFIT}.

\subsubsection*{Noradrenaline constant source $c_n$}
We fund in \ref{app-NeuroInitialValues} that in normal conditions (i.e. in the absence of a tumor) the level of noradrenaline is $N_n^{eq}= 300 \mbox{ pg/mm}^3$. We can then calculate $c_n$ from the equilibrium equation
$$
c_n + s_n S^{eq} - d_n N_n^{eq} = 0 \quad \Rightarrow \quad c_n \approx 500 \; \frac{\mbox{pg}}{\mbox{mm}^3\mbox{day}} \; ,
$$
where $S^{eq}$ and $P^{eq}$ were also found in \ref{app-NeuroInitialValues} and $s_n$,$d_n$ were estimated above.


\subsection{$N_a$-equation}

\subsubsection*{Acetylcholine release rate $s_a$}

In \cite{paton1971} the output of acetylcholine from the plexus of the guinea-pig ileum longitudinal strip is used to study the mechanism of acetylcholine
release. The resting output is reasonably constant for a given preparation for long periods; the mean value for eighty-four experiments was about 51 ng/g$\cdot$min. The evoked output, however, usually changes as stimulation is prolonged, in a manner varying with the stimulation used.
Assuming a nerve cell volume of $10^{-5}\mbox{ mm}^3$ (see \ref{app-NeuroSizeWeight}) and of density equal to water's one (1g/mL), we have that 1 g of parasympathetic nerves corresponds to approximately $10^8$ cells.
Therefore, we estimate the acetylcholine production rate as $s_a = 0.73 \,\nicefrac{\mbox{pg}}{\mbox{cell day}}$.

\subsubsection*{Acetylcholine decay rate $d_a$}
In \cite{bechem1981} the authors  studied the influence of the stimulus interval and the effect of Mn ions on facilitation\footnote{\textsc{Note}: Here the term \emph{facilitation} denotes an increase in transmitter release during repetitive nerve excitation} of acetylcholine (ACh) release from parasympathetic nerve terminals in quiscent guinea-pig auricles.
Here we also find that when conditioning trains of stimuli were applied, a second much longer lasting component of facilitation was found ($t_{1/2}\approx 4$ s).  Also, the decay to the control level displays a half time of about 20 min and can also be accelerated by frequent stimulation of the parasympathetic nerve fibres.
In this regard we can estimate $d_a = 49.91 \mbox{day}^{-1}$ (taking 20 min).

\subsubsection*{Acetylcholine constant source $c_a$}
In \ref{app-NeuroInitialValues} we estimated that in normal conditions (i.e. in the absence of a tumor) the acetylcholine level in the tissue is $N_a^{eq}= 530 \mbox{ pg/mm}^3$. We can then calculate $c_a$ from the equilibrium equation
$$
c_a + s_a P^{eq} - d_a N_a^{eq} = 0 \quad \Rightarrow \quad c_a \approx 26.5\times 10^3  \, \frac{\mbox{pg}}{\mbox{mm}^3\mbox{day}} \; ,
$$
where $S^{eq}$ and $P^{eq}$ were also found in \ref{app-NeuroInitialValues} and $s_a$,$d_a$ were estimated above.

\bibliography{mybib-neuro,mybib-neuro-main}
\bibliographystyle{amsedit}




\newpage 

\tableofcontents

\end{document}